\documentclass[review]{elsarticle}

\usepackage{lineno,hyperref}
\modulolinenumbers[5]

\journal{Elsevier Computer Communications}

\usepackage{tabularx}
\usepackage{booktabs}
\usepackage{placeins}
\usepackage{graphicx}
\usepackage{color}
\usepackage{colortbl}
\usepackage{xcolor}

\usepackage[nonumberlist]{glossaries}

\newacronym{qoe}{QoE}{Quality of Experience}
\newacronym{psnr}{PSNR}{Peak Signal to Noise Ratio}
\newacronym{vr}{VR}{Virtual Reality}
\newacronym{svc}{SVC}{Scalable Video Coding}
\newacronym{svr}{SVR}{Support Vector Regression}
\newacronym{ar}{AR}{Augmented Reality}
\newacronym{ssim}{SSIM}{Structural Similarity Index}
\newacronym{vifp}{VIFP}{Visual Information Fidelity in Pixel Domain}
\newacronym{mos}{MOS}{Mean Opinion Score}
\newacronym[\glslongpluralkey={Fields of View}]{fov}{FoV}{Field of View}
\newacronym[\glslongpluralkey={Groups of Pictures}]{gop}{GoP}{Group of Picture}
\newacronym{qac}{QAC}{Quality-Aware Clustering}
\newacronym{fsim}{FSIM}{Feature Similarity Index}
\newacronym{jvet}{JVET}{Joint Video Exploration Team}
\newacronym{mse}{MSE}{Mean Square Error}
\newacronym{dash}{DASH}{Dynamic Adaptive Streaming over HTTP}
\newacronym{cnn}{CNN}{Convolutional Neural Network}
\newacronym{rnn}{RNN}{Recurrent Neural Network}
\newacronym{srd}{SRD}{Spatial Representation Description}
\newacronym{scp}{SCP}{Shared Coded Picture}
\newacronym{fsm}{FSM}{Fused Saliency Map}
\newacronym{acr}{ACR}{Absolute Category Rating}
\newacronym{mec}{MEC}{Mobile Edge Computing}
\newacronym{dcr}{DCR}{Degradation Category Rating}
\newacronym{gan}{GAN}{Generative Adversarial Network}
\newacronym{gvbs}{GBVS}{Graph-Based Visual Saliency}
\newacronym{itu}{ITU}{International Telecommunication Union}
\newacronym{avc}{AVC}{Advanced Video Coding}
\newacronym{bms}{BMS}{Boolean Map Saliency}
\newacronym{omaf}{OMAF}{Omnidirectional Media Format}
\newacronym{sssim}{S-SSIM}{Spherical SSIM}
\newacronym{spsnr}{S-PSNR}{Sphere-based PSNR}
\newacronym{wspsnr}{WS-PSNR}{Weighted to Spherically Uniform PSNR}
\newacronym{wsssim}{WS-SSIM}{Weighted to Spherically Uniform SSIM}
\newacronym{vvc}{VVC}{Versatile Video Coding}
\newacronym{cpppsnr}{CPP-PSNR}{PSNR for Craster Parabolic Projection}
\newacronym{msssim}{MS-SSIM}{Multiscale SSIM}
\newacronym{pvq}{PVQ}{Perceptual Video Quality}
\newacronym{drl}{DRL}{Deep Reinforcement Learning}
\newacronym{nqq}{NQQ}{Normalized Quality versus Quality factor}
\newacronym{hevc}{HEVC}{High Efficiency Video Coding}
\newacronym{bp}{BP}{Back Propagation}
\newacronym{qavr}{QAVR}{Quality Assessment in VR systems}
\newacronym{qsfs}{QSFS}{Quality Score Fusion Strategy}
\newacronym{cppsnr}{CP-PSNR}{Content Preference PSNR}
\newacronym{cpssim}{CP-SSIM}{Content Preference SSIM}
\newacronym{deepvriqa}{DeepVR-IQA}{Deep VR Image Quality Assessment}
\newacronym{hmd}{HMD}{Head-Mounted Display}
\newacronym{niqe}{NIQE}{Natural Image Quality Evaluator}
\newacronym{sisblim}{SISBLIM}{Six-Step Blind Metric}
\newacronym{cdn}{CDN}{Content Delivery Network}
\newacronym{lstm}{LSTM}{Long Short-Term Memory}
\newacronym{av1}{AV1}{AOMedia Video 1}
\newacronym[\glslongpluralkey={Radio Access Technologies}]{rat}{RAT}{Radio Access Technology}
\newacronym{mdp}{MDP}{Markov Decision Problem}
\newacronym{ocp}{OCP}{Offset Cubic Projection}
\newacronym{tsp}{TSP}{Truncated Square Pyramid}
\newacronym{dmos}{DMOS}{Differential Mean Opinion Score}
\newacronym{rsp}{RSP}{Rotated Sphere Projection}
\newacronym{erp}{ERP}{Equirectangular Projection}
\newacronym{cmp}{CMP}{Cubic Mapping Projection}
\newacronym{vcnn}{V-CNN}{Viewport-based CNN}
\newacronym{mc360}{MC360IQA}{Multi Channel 360$^\circ$ Image Quality Assessment}
\newacronym{dsis}{DSIS}{Double Stimulus Impairment Scale}
\newacronym{rbm}{RBM}{Rhombic Mapping}
\newacronym{npcm}{NPCM}{Nested Polygonal Chain Mapping}
\newacronym{sp}{SP}{Sinusoidal Projection}
\newacronym{opv}{OPV}{Optimal Probabilistic Viewport}
\newacronym{dct}{DCT}{Discrete Cosine Transform}
\newacronym{qp}{QP}{Quantization Parameter}
\newacronym{sao}{SAO}{Sample Adaptive Offset}
\newacronym{knn}{k-NN}{k-Nearest Neighbors}

\makeglossaries

\begin{document}

\begin{frontmatter}

\title{A Survey on 360$^\circ$ Video: Coding, Quality of Experience and Streaming}

\author{Federico Chiariotti$^*$}
\address{Aalborg University}
\address{Fredrik Bajers Vej 7C, 9220 Aalborg, Denmark}
\cortext[mycorrespondingauthor]{Corresponding author}
\ead{fchi@es.aau.dk}

\begin{abstract}
The commercialization of \gls{vr} headsets has made immersive and 360$^{\circ}$ video streaming the subject of intense interest in the industry and research communities. While the basic principles of video streaming are the same, immersive video presents a set of specific challenges that need to be addressed. In this survey, we present the latest developments in the relevant literature on four of the most important ones: \emph{(i)} omnidirectional video coding and compression, \emph{(ii)} subjective and objective \gls{qoe} and the factors that can affect it, \emph{(iii)} saliency measurement and \gls{fov} prediction, and \emph{(iv)} the adaptive streaming of immersive 360$^\circ$ videos. The final objective of the survey is to provide an overview of the research on all the elements of an immersive video streaming system, giving the reader an understanding of their interplay and performance.
\end{abstract}


\end{frontmatter}

\glsresetall


\section{Introduction}\label{sec:intro}
Over the past few years, the commercialization of \gls{vr} headsets and cheaper systems using smartphones as viewports~\cite{amin2016immersion} have fueled a strong research interest in 360$^{\circ}$ immersive videos, and the technology is currently undergoing standardization~\cite{skupin2017standardization}. Commercial \glspl{hmd} are currently being sold by multiple companies, and the artistic potential of the new medium is being explored for both gaming and movies.

This kind of technology has the potential to make video a more intense experience, with a stronger emotional impact~\cite{visch2010emotional}, thanks to the wider \gls{fov} and the direct user control of viewing direction. 360$^{\circ}$ videos also have a huge potential for storytelling, as multiple story lines can be developed in parallel~\cite{lescop2017narrative}. Immersive video also has the potential to enhance empathy and participation in news stories~\cite{de2010immersive}, although evidence regarding its use shows mixed results~\cite{wang2018effects}. Psychological factors such as perception of embodiment~\cite{schultze2010embodiment} also affect immersiveness~\cite{steed2016in}, particularly when an avatar is animated in the \gls{vr} simulation~\cite{lin2013stepping}. 

Immersive video streaming presents some unique challenges~\cite{zink2019scalable}, especially for live streaming: since the full omnidirectional view is wider than traditional video, it requires far more bandwidth to be streamed with a comparable quality. In order to reduce the throughput of 360$^\circ$ streams~\cite{afzal2017characterization}, tile-based solutions have become a standard: the sphere is divided in several tiles, according to a pre-defined projection scheme, and each tile can be downloaded as a separate object. In this way, clients can concentrate most of their resources on the tiles that are in the user's \gls{fov}, i.e., the ones that will actually be displayed with the highest probability, resulting in the same \gls{qoe} even if tiles outside the viewport have a very low resolution or are not downloaded at all. Naturally, this kind of solution requires an accurate prediction of where the user's gaze will fall, which is in itself a complex research topic. The design of the tiling scheme is also a significant factor in both the compression efficiency of the video coding scheme and the final \gls{qoe} of the user.

Additionally, the geometric distortion~\cite{li2017spherical} generated by the projection of spherical omnidirectional video onto a flat surface reduces both the accuracy of traditional \gls{qoe} metrics and the efficiency of 2D video codecs. 
Since traditional \gls{qoe} metrics are designed for planar images and videos, their direct use does not correctly represent the human perception of the video and is only loosely correlated with actual \gls{qoe}. The design of projective corrections for legacy metrics and 360-specific ones is an active area of research. Cybersickness~\cite{kim2019vrsa} is another major problem for immersive video streaming, requiring both a more precise metric to evaluate quality variations and better streaming techniques to reduce stalling.

The distortion issue also affects automatic saliency estimation, which can help predict the \gls{fov}, and even feature extraction and \glspl{cnn}~\cite{yu2015framework} are affected by it, requiring \emph{ad hoc} corrective methods~\cite{su2017learning}.

This survey aims at providing readers with a broad overview of the state of the art in all the major research directions on omnidirectional video. We give a full perspective on the building blocks of an omnidirectional streaming system: 
\begin{itemize}
 \item In Sec.~\ref{sec:coding}, we examine coding methods, discussing different standards and projections and how they can introduce different kinds of distortion and enable more efficient compression;
 \item In Sec.~\ref{sec:quality}, we describe subjective and objective metrics to evaluate the \gls{qoe} of omnidirectional videos, and why it is a complex challenge;
 \item In Sec.~\ref{sec:saliency}, we examine the question of saliency and \gls{fov} prediction. We review empirical approaches based on user behavior, analytical ones based on image features, and joint ones that consider both past viewport directions and the current image;
 \item In Sec.~\ref{sec:streaming}, we present the state of the art on omnidirectional video streaming techniques, focusing on tiling-based approaches. We also review some recent network-level innovations to provide support to omnidirectional streaming.
 \item In Sec.~\ref{sec:concl}, we present a summarized version of the lessons learned on each topic and conclude the paper with a discussion of the open research challenges in the field.
\end{itemize}
Each section of the paper includes a discussion of the key challenges and open problems in the field.

A number of recent surveys, whose contribution is summarized in Table~\ref{tab:surveys}, have examined the state of the art on different topics in the field.
One work~\cite{chen2018recent} focuses on projection, explaining several state of the art methods in detail and evaluating them on a public dataset with known quality metrics. The authors explore viewport-adaptive coding as a possible solution to the demanding bandwidth requirements of omnidirectional video, and briefly mention the possible sources of coding distortion, which are examined in detail in~\cite{azevedo2019visual}: this work considers the steps of the encoding chain, examining how each one introduces different kinds of local and global image distortions. 
A more recent work~\cite{xu2020state} takes a broader view, examining the existing \gls{qoe} evaluation metrics, along with viewer attention models for eye and head movements, while the networking aspects of streaming, from resource allocation to caching, are reviewed in~\cite{he2018network}.
Finally, a survey focusing on system design and implementation~\cite{fan2019survey} examines some of the existing systems, protocol and standards for acquisition, compression, transmission, and display of omnidirectional videos. 

These recent works only present a limited review of \gls{fov}-adaptive streaming, while our Sec.~\ref{sec:streaming} has a more extensive review of the existing literature. Furthermore, while all of these works concern themselves with \gls{qoe}, this work is the first to provide an analysis of the existing comparisons between objective metrics, resulting in insights for further research and implementation. Finally, these recent surveys only present a limited review of \gls{fov}-adaptive streaming, and none of them has a complete perspective that unifies evaluation and streaming: since the efficiency of adaptation techniques strongly depends on both the encoding techniques and the \gls{fov}, presenting them in a unified manner is important to get a full picture of the design requirements. The discussion of the field developed in this survey has a unified perspective, linking the later sections to the earlier ones and proposing some ideas for a holistic development of 360$^\circ$ streaming systems.

\begin{table}[t]
	\centering 
	\caption{Summary of the existing surveys on omnidirectional video}
	\label{tab:surveys}
	\tiny
	\begin{tabular}{p{7cm}p{0.5cm}p{3cm}}
		\toprule
		Survey & Year & Topic  \\
		\midrule
		Recent advances in omnidirectional video coding for virtual reality: Projection and evaluation~\cite{chen2018recent} & 2018 & Projection\\
    Visual Distortions in 360-degree Videos~\cite{azevedo2019visual} & 2019 & Visual distortion\\
    State-of-the-Art in 360$^\circ$ Video/Image Processing: Perception, Assessment and Compression~\cite{xu2020state} & 2020 & Saliency, \gls{qoe}\\
    Network Support for AR/VR and Immersive Video Application: A Survey~\cite{he2018network} & 2018 & System implementation\\
    A Survey on 360$^\circ$ Video Streaming: Acquisition, Transmission, and Display~\cite{fan2019survey} & 2019 & Protocol design\\
		\bottomrule
	\end{tabular}
\end{table}

\section{Coding, compression and distortion}\label{sec:coding}

The efficient encoding of omnidirectional video has all the well-known issues of 2D video encoding, with an additional degree of complexity: since filters and coding tools are often based on 2D images, the spherical content needs to be projected to a flat surface to be processed and encoded. In this section, we discuss the different factors that should be considered when encoding omnidirectional video, presenting the main projection schemes and coding solutions, both in the spatial and temporal domains.

\subsection{Projection and tiling}\label{ssec:tiling}

The geometric distortion issue in 360$^{\circ}$ video is the same that cartographers have faced for thousands of years when drawing maps of the Earth~\cite{snyder1997flattening}: projecting a sphere onto a planar surface inevitably leads to some form of distortion. However, projection is not the only source of distortion, as the omnidirectional video processing pipeline can cause it at every step~\cite{azevedo2019visual}. The first one is the acquisition of the image: omnidirectional images and videos are usually stitched from multiple cameras~\cite{szeliski2007image}, and this can introduce several kinds of issues at the edges. These can range from missing information and misalignment of the edges to differences in the exposure and ``ghosting'', and are often particularly strong at the poles, which most camera systems cannot capture and are often reconstructed in post-processing. Video can also have temporal discontinuities, such as objects appearing and disappearing or warping as objects move close to the stitching areas~\cite{jiang2015video}. In order to avoid smoothness issues and increase the coding efficiency, appropriate motion models that explicitly use rotation need to be used~\cite{vishwanath2017rotational}.

After the omnidirectional image has been acquired, it needs to be converted to a planar representation for encoding and storage. It can then be divided into tiles to allow tile-based streaming, which we will discuss in detail in Sec.~\ref{sec:streaming}. The warping patterns generated by the combination of the map projection and tile edges will then interact. Consequently, the form and severity of the geometric distortion effects depend strongly on the projection and tiling scheme, which is crucial for efficient compression of omnidirectional video.

The \gls{erp}~\cite{salomon2007transformations} is the oldest, simplest, and most common projection for omnidirectional video: it is similar to the \emph{plate carr\'{e}e} geographic projection, as it divides the sphere of view in a number of rectangles with the same solid angle. Distortion at the poles makes projection wasteful, as it encodes the poles with more pixels than the equator: as viewers usually focus their attention close to the equator, the poles are often outside the \gls{fov}. 

The dyadic projection~\cite{benko2014dyadic} tries to solve the pole oversampling issue by reducing the sampling for vertical angles above $\frac{\pi}{3}$ from the equator, while the barrel projection~\cite{youvalari2016efficient} encodes the top and bottom quarters of the \gls{erp} as circles, reducing the number of pixels used for the two caps. The polar square projection~\cite{wang2017polar,jallouli2019equatorial} is another adaptation that works like the barrel projection, but maps the poles to two squares. There are other techniques to compensate for the pole oversampling issue: the equal-area cylindrical projection~\cite{safari2007new} reduces the height of the tiles with the latitude, while the latitude adaptive approach~\cite{lee2017omnidirectional} adapts the number of tiles to the latitude. The result is also known as \gls{rbm}~\cite{wu2017rhombic}, since the tiles are arranged in a rhombic shape, which can then be rearranged onto a rectangle. The octagonal projection~\cite{chengjia2018octagonal} does the same with a rough latitude quantization, resulting in its namesake shape. \gls{npcm} is another downsampling technique~\cite{kammachi2017nested}, which starts from the \gls{erp} output and linearly approximates the optimal sampling density.

\begin{figure}
\centering
\includegraphics[width=3.4in]{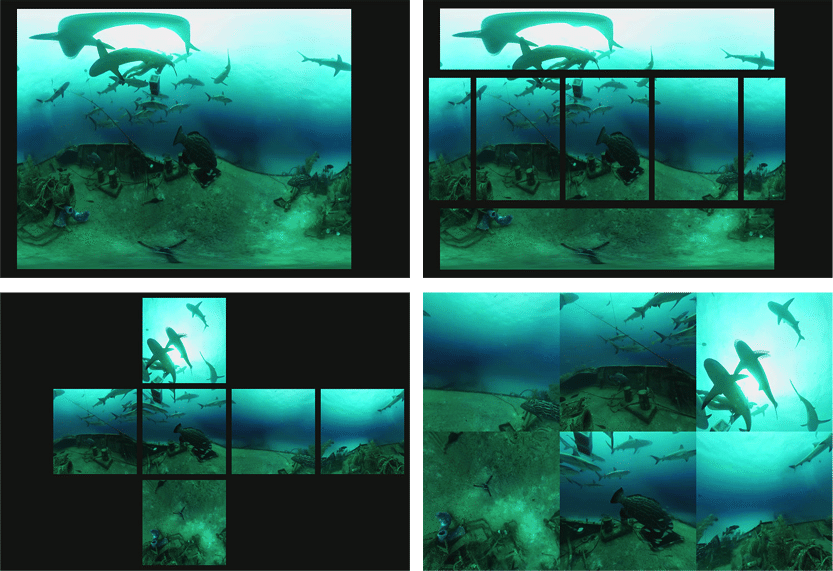}
\caption{Equirectangular and cubemap projection comparison. The figure was adapted from the Facebook video engineering blog: \url{https://engineering.fb.com/video-engineering/under-the-hood-building-360-video/}}
\label{fig:erp}
\end{figure}

The \gls{cmp} is the other projection to be widely adopted. It constructs a cube around the sphere~\cite{li2017projection}, then projects rays outward from the center. Each ray intersects with a single point on the surfaces of both solids, resulting in the projection mapping. The \gls{cmp}~\cite{gomez2018ticmp} is more efficient than the \gls{erp} in terms of compression~\cite{zhou2016study}, and is currently used by Facebook for omnidirectional videos~\cite{zhou2017measurement}. A comparison between the \gls{erp} and \gls{cmp} projections is shown in Fig.~\ref{fig:erp}. It is easy to see that distortion at the poles is far lower, while objects close to the edges and corners of a face are more distorted. This should be intuitive, as the cube mapping approximates a sphere better close to the center of each face: this effect can be mitigated by applying equiangular mapping to the cube faces~\cite{lin2019efficient}, or in general by adjusting the sampling to privilege the center of each face~\cite{he2018content}.

\begin{table}
	\centering 
	\caption{Summary of state of the art projections}
	\label{tab:projections}
	\tiny
	\begin{tabular}{p{2.5cm}p{4cm}p{3.5cm}}
		\toprule
		Projection & Geometry & Main advantages and issues\\
		\midrule
		Equirectangular~\cite{salomon2007transformations} & Each rectangle has the same solid angle & Oversampling at the poles\\
		Dyadic~\cite{benko2014dyadic} & Equirectangular with reduced polar sampling & Distortion at the poles\\
		Barrel~\cite{youvalari2016efficient} & The sphere is mapped to a cylinder & Distortion at the edges\\
		Polar square~\cite{wang2017polar} & Barrel-like, mapping the poles to squares & Distortion at the poles\\
		Equal-area cylindrical~\cite{safari2007new}& Equirectangular with latitude-dependent tile height & Reduced polar oversampling\\
		Latitude adaptive~\cite{lee2017omnidirectional} & Equirectangular with latitude-dependent number of tiles & Reduced polar oversampling\\
		Rhombic mapping~\cite{wu2017rhombic} & Similar to latitude adaptive, arranging tiles in a rhombus & Efficient retiling\\
		Nested polygonal chain~\cite{kammachi2017nested} & Downsampling from equirectangular & Reduced polar oversampling\\
		Cubic mapping~\cite{li2017projection} & Projection from sphere to cube & Higher efficiency, lower polar distortion, edge distortion\\
		Equiangular cubic mapping~\cite{lin2019efficient} & Equiangular mapping on cube faces & Reduced face edge distortion\\
		Other solids~\cite{lin2016ahg,fu2009rhombic,akula2017ahg} & Projection on solids with more faces & Lower projection distortion, higher edge distortion\\
		Variable tile shape~\cite{li2016novel} & Tiles can be adapted to the content & Low distortion, complex encoding and decoding\\
		Rotated sphere~\cite{abbas2017ahg} & Baseball-like unfolding & Increased coding efficiency, low edge distortion\\
		ClusTile~\cite{zhou2018clustile} & Viewer behavior-based adaptive sampling & Low distortion, complex encoding and decoding\\
		\bottomrule
	\end{tabular}
\end{table}

\begin{figure}
\centering
\includegraphics[width=3.4in]{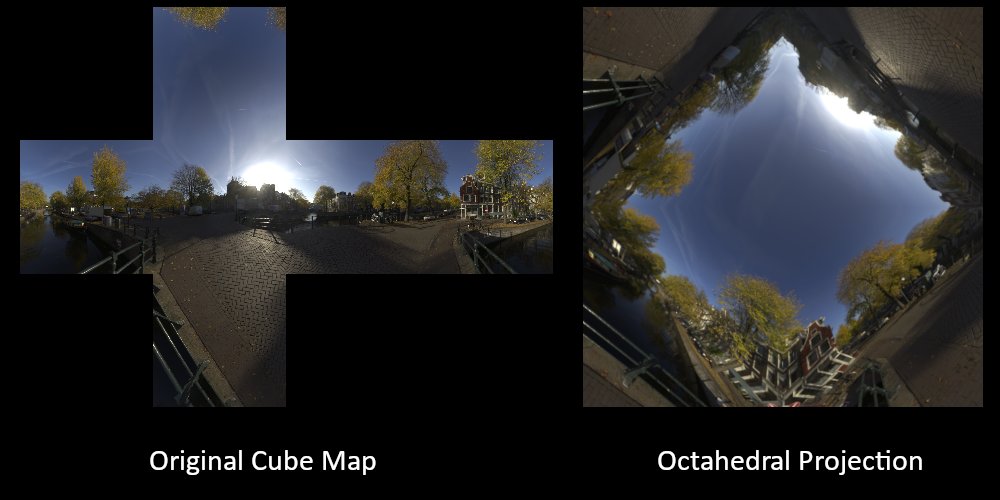}
\caption{Equirectangular and octahedral projection of the same scene. Image credits: Omar Shehata, \url{https://omarshehata.me/}}
\label{fig:oct}
\end{figure}

Solids with a larger number of faces, such as octahedrons~\cite{lin2016ahg}, rhombic dodecahedrons~\cite{fu2009rhombic}, or icosahedrons~\cite{akula2017ahg}, can reduce the effect of edges by having a lower stretch and area distortion, like the \gls{sp}~\cite{seong2002sinusoidal}, which is an equal area projection. However, there is a trade-off when choosing the number of faces: polyhedrons with more faces have a lower projection distortion, but a higher number of discontinuous boundaries. An example of octahedral projection is shown in Fig.~\ref{fig:oct}. Other less regular projection shapes are also possible, with tiles of variable size and shape~\cite{li2016novel}. The \gls{rsp}~\cite{abbas2017ahg} unfolds the sphere under two rotation angles and stitches them like a baseball; this can be obtained from the \gls{erp}, and it can increase coding efficiency. 

Finally, a more advanced approach to projection integrates content and viewer behavior in the design~\cite{yu2015content}: areas that have salient content and are often watched will be sampled at a higher rate. ClusTile~\cite{zhou2018clustile} is another projection that uses past viewer behavior, designing a set of tiles that minimizes bandwidth requirements for past views. A framework evaluating the projections presented above was described in~\cite{yu2015framework}, and some results comparing the basic projections' compression efficiency and distortion with H.264 and H.265 codecs are presented in~\cite{li2016evaluation}, finding that the equal-area cylindrical projection outperforms both the \gls{erp} and \gls{cmp}.
The main projection methods we presented in this section are summarized in Table~\ref{tab:projections}.

\begin{figure}
\centering
\includegraphics[width=3.4in]{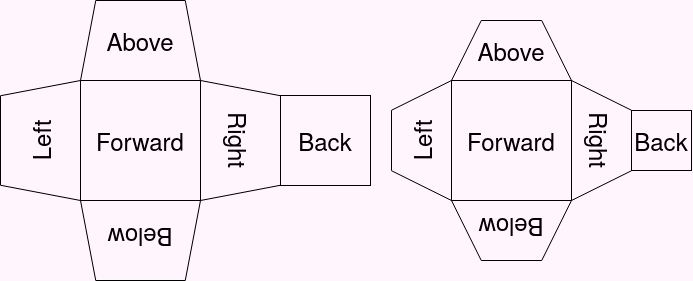}
\caption{Truncated square pyramid projection with different settings.}
\label{fig:tsp}
\end{figure}

Offset projection is a concept meant to save bandwidth and exploit the available knowledge of the user's viewing direction: offset projections use more pixels to encode regions close to the predicted gaze direction, while regions at wide angles from it have a higher compression. The \gls{tsp}~\cite{van2017ahg} projection constructs a truncated pyramid around the sphere, with the bottom facing the same way as the viewer. The projection is then constructed like the \gls{cmp}. The construction of the solid is shown in Fig.~\ref{fig:tsp}, in which two truncated pyramids with different settings are shown: the one on the right has a smaller upper base, giving more relative importance and more pixels to the region facing the viewport directly.
When the pyramid's upper base is very small, regions at wide angles from the user's expected gaze are encoded by very few pixels~\cite{zare2017virtual}, with extreme compression gains. 

The \gls{ocp}~\cite{zhou2018effectiveness} adopts another way to perform offset projection: it is a version of the \gls{cmp}, with an offset that distorts the sphere before projecting it to the six cube faces. In the resulting frame, views in one direction have a higher pixel density than in other directions. The same concept can be applied to any combination of the equirectangular and barrel projection, and a possible option is to consider only an offset on the horizontal plane. Offset projections can significantly improve the \gls{qoe} of an omnidirectional image, as long as the view orientation is close to the offset. Another offset projection is the asymmetric circular projection~\cite{wang2017asymmetric}, which decreases sampling density in the area outside the \gls{fov} smoothly by using a circle with a center closer to the surface in the direction of the user's gaze. In this way, there are no explicit seams. If an \gls{fov} prediction is available, streaming clients can select the appropriate offset orientation and increase \gls{qoe} without a corresponding throughput increase~\cite{chen2018recent}. The same operation can be performed for the equirectangular and barrel projection. An evaluation of the quality of different offset projections is available in~\cite{zhou2018effectiveness}, for different viewing angle errors and offset distortion settings. 

\subsection{Compression and encoding}\label{ssec:encoding}

There are a number of competing video encoding standards being developed~\cite{grois2016coding}: the most popular are \gls{hevc}~\cite{pourazad2012hevc}, or H.265, and \gls{av1}~\cite{chen2018overview}, but the older \gls{avc}~\cite{bauermann2006h}, or H.264, is still widely used. Additionally, \gls{vvc}~\cite{ye2019omnidirectional}, the future H.266, promises to add new capabilities to the existing standards. The 2D encoding techniques in the standards are highly optimized and close to ubiquitous, and most omnidirectional streaming systems reuse the 2D coding pipelines~\cite{zare2016hevc}. However, all the distortion issues discussed in Sec.~\ref{ssec:tiling} do not just impact the \gls{qoe} of the projected and encoded video, but also the coding efficiency. Furthermore, the resampling and interpolation steps of the encoding pipeline often cause aliasing and blurring, and if these steps are not managed carefully~\cite{bagnato2012plenoptic} they can also introduce visible seams and combine with the projection scheme to create distortion. While older works can get good results using custom techniques on the spherical image, often without projection~\cite{tosic2009low}, most of the recent literature follows the standard approach, with all its advantages and pitfalls. The decision on the representations that need to be encoded and stored~\cite{ozcinar2017estimation} in a streaming system can affect the requirements on bandwidth support, server storage space and distortion.

Naturally, coding efficiency depends on the projection used, and it is possible to optimize coding for a certain projection, reducing its downsides and increasing compression performance.
Since \gls{erp} oversamples the polar regions, it is possible to use smoothing~\cite{budagavi2015360} or reduce the accuracy of motion vectors and the coding block resolution~\cite{ray2018low} as a function of the latitude, increasing the coding efficiency with minimal \gls{qoe} impacts. Another way to compensate for this distortion is to adaptively set the \glspl{qp}, using the \gls{wspsnr} weights: regions that are less important in the metric will be encoded with a rougher compression~\cite{li2017spherical}.  The same optimization can be performed for other metrics, such as \gls{spsnr}~\cite{liu2018rate}. A more advanced way to set the \glspl{qp} is to combine the geometric information with the saliency~\cite{luz2017saliency}, privileging the salient areas which will be watched more often.

The \gls{erp} latitude-adaptive quantization technique is adopted in~\cite{zhang2019efficient}, combined with some steps to terminate the coding unit partition early in these areas, speeding up the encoding process. Early coding unit termination can also be performed in a content-dependent way~\cite{zhang2019fast}, computing the local texture complexity. Another optimization for \gls{erp} concerns the edges of the image: since the left and right edges are actually continuous, the coding unit parameters need to be set to avoid visible seams~\cite{li2018reference}. In~\cite{tang2017optimized}, the region-adaptive quantization scheme is combined with an adaptive mechanism that reduces the frame rate to increase picture quality if the motion in the content is not too fast. An alternative strategy is rotation: since regions close to the equator have less distortion, interesting regions of the image with high motion and fine-grained textures can be rotated to the equator, while the less interesting regions are rotated to the poles and have more distortion~\cite{boyce2017spherical}. This approach is extended in~\cite{su2018learning}, using a \gls{cnn} to predict the orientation that maximizes the achievable compression over the \gls{gop}, as both content and motion vector discontinuities can affect the compressibility.

Filters are another important concern in omnidirectional coding, as their effectiveness relies on proper adaptation to the projection. In~\cite{zhou2019fast}, the \gls{sao} filter that can improve coding quality for sharp edges is adapted to the \gls{erp}, reducing the coding complexity by up to 80\% with no \gls{qoe} impacts. A correction to the standard \gls{hevc} deblocking filter can reduce the \gls{cmp} edge distortion~\cite{sauer2018geometry} by aligning the face edges with the filter edges, filtering only the left and top borders to maintain rotational symmetry, and using the correct pixels in the 3D representation for the filter decision-making. A similar approach is used in~\cite{lin2019efficient}, limiting the coding unit splits at the face edges and adapting the \gls{hevc} filter to the equiangular \gls{cmp} by enforcing the face boundaries and using the correct pixels. The authors also adapt a \gls{cnn} denoising filter to the projection. Coding unit depths can also be adapted to the content and \gls{cmp} geometry, reducing coding time significantly~\cite{guan2019fast}. 

In projections with more irregular face shapes, the inactive samples that are used to pad the 2D projected frame to a rectangular shape can be ignored in the rate-distortion optimization, resulting in further compression benefits~\cite{herglotz2019efficient}. A full coding system using a sampling-adjusted \gls{cmp} is presented in~\cite{hanhart2018360b}, including padding and other techniques to limit face boundary discontinuities such as packing, i.e., reshuffling of the cube faces in the representation so that contiguous objects in the 3D sphere are close in the projected image.

\subsection{Motion estimation and temporal coding}

The temporal element is critical when encoding omnidirectional video: since the content is dynamic and encoded in \glspl{gop}, considering the motion in subsequent frames significantly increases the compression efficiency. The first example is downsampling: performing the operation on each frame statically does not achieve the same compression efficiency as considering the quality of the dependent B and P frames~\cite{youvalari2016analysis} when downsampling the independent I frames they are tied to. It is also possible to reduce the number of independent frames by adopting the \gls{scp} technique, which introduces P-coded pictures that are the same across all representations. This enables longer \glspl{gop}, increasing the efficiency of the code, but also the encoding and decoding complexity~\cite{ghaznavi2018shared}.

Motion estimation is inextricably tied into saliency, which we will discuss in Sec.~\ref{sec:saliency}: the content that is most important to viewers, and on which their gaze usually fixates, is often also the fastest-moving one. This has important consequences for streaming systems which use prediction of the future \gls{fov} to optimize the bandwidth utilization, as these systems require accurate predictions and efficient coding. As the use of offset projection, temporal coding, and \gls{fov}-oriented predictive streaming all aim at improving compression while maintaining an accurate representation of moving content, the interplay of these subsystems must be considered when designing a streaming system.

The effects of projection also complicate motion modeling in omnidirectional video: since projection is a non-linear transformation, a simple translational motion of all the projected pixels in a local block (like in the \gls{hevc} standard) will not be able to capture the actual motion of the content. This distortion can become catastrophic if the motion crosses face boundaries, causing texture discontinuities that seriously impair \gls{qoe}.

A possible solution is to reproject the motion vector: if the motion on the sphere is translational (i.e., the movement is on the surface of the sphere), the motion vector on the projected video is converted to the spherical motion vector, which is then interpolated~\cite{de2017deformable}. In this way, the coding efficiency and the \gls{qoe} increase; the same can be done for purely rotational motion. This technique was proposed for the \gls{cmp}~\cite{li2017projection,lin2019efficient} and \gls{erp}~\cite{vishwanath2017rotational}, integrating it with standard \gls{hevc} motion modeling schemes. In~\cite{li2018advanced}, a general model is tested for \gls{erp}, \gls{cmp} and octahedron projections. The spherical coordinate transform can be used to further improve performance and extend the possible motions to the whole 3D space~\cite{wang2019spherical}, working in spherical coordinates and using relative depth to convert between \gls{erp} and the 3D space. It is also possible to assign different motion vectors to pixels in the same block, correcting the motion vector distortion~\cite{zheng2007adaptive}. A less efficient but less computationally demanding way to correct motion vectors in \gls{erp} is to exploit the \gls{wspsnr}~\cite{sun2016ahg} weight map to calculate a scaling factor for the motion vectors~\cite{ghaznavi2018geometry}. 

Another technique deals with distortion due to motion compensation failures at face boundaries extends a face by linearly projecting the pixels in the other faces~\cite{he2016geometry} to preserve texture continuity~\cite{ma2016coprojection}. This operation can be performed more efficiently using polytope geometry~\cite{sauer2017improved}. Another work~\cite{li2018advanced} considers the angle of the block in the sphere in the \gls{erp} projection when computing the padding.

Deep learning is a new alternative to traditional motion estimation: in~\cite{li2018convolutional}, \glspl{cnn} are used to reconstruct future cubemap frames, combining the encoded P or B frame with the last received I frame. This scheme can improve \gls{psnr} without increasing the required bandwidth.

\section{Quality of Experience in Immersive Videos}\label{sec:quality}

\gls{qoe} is the ultimate measure of performance for both standard and panoramic video streaming. However, its subjective nature makes finding a general metric to measure it extremely difficult~\cite{skorin2018survey}. Although most of the research on standard video is still applicable, 360$^{\circ}$ video presents some unique challenges~\cite{perrin2017measuring}: an important factor in the perceived quality of panoramic video is the geometric distortion given by the projection of the spherical image on a planar display~\cite{jabar2017perceptual}, which is more pronounced with wide \glspl{fov}. It is possible to assess these distortions objectively~\cite{jabar2018objective}, but not their impact on \gls{qoe}. For a more comprehensive survey on the possible sources of distortion in 360$^{\circ}$ videos, we refer the reader to~\cite{azevedo2019visual}. 
Another important factor in the quality of omnidirectional video is the mosaic technique, which can generate distortion in dynamic scenes~\cite{gurrieri2013acquisition}. 

In this section, we consider subjective and objective methods to measure omnidirectional video \gls{qoe}, and present the wide body of literature on the evaluation of these metrics. We conclude the section with a discussion of dynamic effects on omnidirectional video \gls{qoe}.

\subsection{Measuring QoE: subjective methods}\label{ssec:qoe_subjective}
\gls{qoe} is a complex concept, as it involves the human interaction with the content, and its automatic assessment is a challenging problem~\cite{akhtar2019multimedia}. Since a direct measure of \gls{qoe} requires human subjects, the assessments need to be performed in controlled and replicable conditions. The standard methodologies for conducting these assessments are specified by the \gls{itu} in~\cite{itu1999subjective}, and distinguish between \gls{acr} and \gls{dcr} scoring. The standard methodologies were developed for 2D video, and they often have to be adapted for omnidirectional video: in~\cite{singla2017comparison}, an example of a new \gls{acr} methodology for omnidirectional video without requiring users to take off their \gls{hmd} is presented. The standard testing conditions specified by the \gls{jvet}~\cite{alshina2017jvet} are also often used, although slightly different from the \gls{itu} recommendations.

The golden standard for \gls{acr} quality assessment is \gls{mos}: the content is shown in controlled conditions to a large number of human subjects, who then rate it on a scale from 1 to 5. When evaluating compression schemes, \gls{dmos} is often used as a \gls{dcr} metric, evaluating the difference between the quality of the compressed content and the original's: this is a fundamental step of the evaluation of new coding schemes, for both standard and omnidirectional content~\cite{xu2018assessing}. Omnidirectional video content is even more challenging, as static image quality is not the only component that influences \gls{qoe}, and even subjective studies need to consider \gls{fov} changes and how the different encoding of foreground and background affects the experience~\cite{curcio2017bandwidth}. A testing methodology that considers the dynamic aspect of \gls{qoe}, accounting for delays between user motion and the high-quality rendering of the video in the new direction, is presented in~\cite{singla2019subjective}.

\gls{dsis} is another way to measure quality impairment of compressed sequences specified in~\cite{itu1999subjective}: instead of rating the content \gls{qoe} on an absolute scale, and possibly comparing it with the unimpaired version's score, this assessment method asks users to rate the degradation directly, after being shown the original and impaired sequence one after the other. However, this method may cause cybersickness more often~\cite{singla2018comparison} when used for omnidirectional video. A more complete comparison between various assessment methods is presented in~\cite{singla2019comparison}.

Immersiveness is another factor that needs to be considered in omnidirectional video \gls{qoe} assessment, as the quality of the video can significantly improve the sense of presence in a \gls{vr} environment. In order to do so, more factors than just picture quality need to be considered~, as audio quality and spatial features can have a strong impact on sense of presence, as well as the proprioceptive matching between the user's movements and the video displayed on the \gls{hmd}~\cite{zou2018framework}. Multi-sensory environments~\cite{wanick2018virtual} that include haptic feedback or even smells present yet more challenges: n~\cite{guedes2019subjective}, immersiveness is evaluated when an external sensory stimulus is combined to the omnidirectional video, finding that this kind of addition can improve immersiveness and enrich user experience.

Finally, an interesting development that straddles the line between subjective and objective metrics is the creation of metrics based on objective physiological data from the user collected by smart watches and other simple sensors~\cite{egan2016evaluation}. In~\cite{arnau2017perceptual}, the authors develop a \gls{qoe} metric based on the combined electroencephalographic, electrocardiographic and electromyographic signals, achieving high correlation with \gls{mos}.

\begin{table}
	\centering 
	\caption{Available subjective \gls{qoe} assessment datasets}
	\label{tab:qoe_datasets}
	\tiny
	\begin{tabular}{lllll}
		\toprule
		Reference & Type & Subjects & Videos or images & Total sequences\\
		\midrule
    \cite{li2018bridge} & Video & 221 & 60 & 600\\
    \cite{xu2017subjective} & Video & 88 & 6 & 48\\
    \cite{singla2017comparison} & Video & 30 & 6 & 60\\
    \cite{sun2018large} & Video & 30 & 13 & 364\\
    \cite{zhang2018subjective} & Video & 30 & 10 & 60\\
    \cite{yang20183d} & Static images & 20 & 16 & 320\\
    \cite{croci2019voronoi} & Video & 21 & 5 & 75\\
    \cite{curcio2017bandwidth} & Video & 12 & 3 & 24\\
    \cite{schatz2017towards} & Video & 27 & 2 & 10\\
    \cite{duan2017ivqad} & Video & 13 & 10 & 150\\
    \cite{zhang2017subjective} & Video & 23 & 16 & 384\\
    \cite{xie2019modeling} & Video & 340 & 30 & 1608 \\
    \cite{yang2018stereoscopic} & Stereoscopic video & 30 & 13 & 364\\
		\bottomrule
	\end{tabular}
\end{table}

Several \gls{qoe} studies have published their datasets, providing a common base for future research on \gls{qoe} assessment. 
The largest dataset is the one presented in~\cite{li2018bridge}, with 221 total subjects watching 60 video sequences, following the methodology described in~\cite{xu2017subjective}, which also presents a public dataset with a total of 88 subjects watching 48 video sequences extracted from 6 videos. The dataset presented in~\cite{singla2017comparison} contains data from 30 users watching 60 sequences, and it was obtained using different methodologies, so it can be used to compare them. In~\cite{sun2018large}, 13 videos are processed into 364 sequences, watched by 30 subjects. In~\cite{zhang2018subjective}, 10 omnidirectional videos of 10 seconds each are evaluated by 30 non-expert subjects. The dataset in~\cite{duan2018perceptual} uses static images, having 20 subjects evaluate 528 compressed versions of 16 base images, as does the one in~\cite{yang20183d}, with 320 compressed versions of 16 images watched by 20 subjects. The authors of~\cite{croci2019voronoi} also released their dataset, with 21 participants watching 75 impaired video sequences with different resolution and compression levels. 
There are other small-scale datasets associated to other measurement studies~\cite{curcio2017bandwidth,schatz2017towards}, while two more large dataset, with 13 subject watching 150 videos and 23 subjects watching 384, were presented in~\cite{duan2017ivqad} and~\cite{zhang2017subjective}, respectively.
To the best of our knowledge, the largest available dataset was presented in~\cite{xie2019modeling}, and is divided in 5 scenarios with an approximately uniform division of samples. Finally, there is a large-scale dataset for stereoscopic omnidirectional video, which was presented in~\cite{yang2018stereoscopic}. The datasets above are summarized in Table~\ref{tab:qoe_datasets}.

\subsection{Objective QoE metrics}\label{ssec:qoe_metrics}

The easiest method to objectively measure the \gls{qoe} of an omnidirectional image is to directly use a classic 2D metric such as \gls{psnr}, \gls{ssim}~\cite{wang2004image}, \gls{msssim}~\cite{wang2003multiscale}, \gls{vifp}~\cite{sheikh2006image}, or \gls{fsim}~\cite{zhang2011fsim}. However, these metrics do not take the geometric distortion caused by the projection of the spherical image into account; indeed, most objective \gls{qoe} metrics for omnidirectional images and videos are adaptations of these metrics, with some corrections for the geometrical distortion resulting from the projection of spherical images on a plane.

\gls{spsnr}~\cite{alshina2017jvet} is an adaptation of \gls{psnr} that takes a number of uniformly distributed sampling points on a spherical surface, then reprojects them on the reference and distorted omnidirectional images and computes \gls{psnr}. Points that are between sampling positions in the 2D plane are mapped to the nearest neighbor. \gls{wspsnr}~\cite{sun2017weighted} takes the opposite approach, computing \gls{psnr} on each pixel of the projected image, then weighting the results proportionally to the area occupied by the pixel on the sphere.
\gls{cpppsnr}~\cite{zakharchenko2016ahg} is a projection-independent adaptation of \gls{psnr}; it applies a Craster parabolic projection that preserves areas in the spherical domain, then calculates \gls{psnr} on the resulting image. By virtue of being independent of the projection used in the image, it allows the comparison of different projection methods.
Finally, \gls{sssim}~\cite{chen2018spherical} and \gls{wsssim}~\cite{zhou2018weighted} are adaptations of \gls{ssim} to the spherical domain: the structural similarity is adjusted to compensate the geometrical distortion using a weighting function similar to the one used by \gls{wspsnr}. In~\cite{croci2019voronoi}, the sphere is divided into patches using a Voronoi diagram, and the 2D algorithms are applied on the patches, reducing the distortion.

The content itself can be the basis of the weighting system, as in~\cite{xu2018assessing}: \gls{cppsnr} and \gls{cpssim} are adaptations of the two metrics that take the viewport direction and content saliency into account, using a predictive model to gauge future viewing direction. However, saliency and eye movement models are not always perfect, and using the center of the viewport as a proxy for gaze direction is still very imprecise~\cite{rai2017which}.

More complex metrics take into account several factors, often combining the objective metrics mentioned above: in~\cite{zou2017perceptual}, a non-linear \gls{pvq} model is derived, starting from \gls{ssim} and other metrics and matching them to a predicted \gls{mos}. The same operation is performed by the \gls{nqq} model in~\cite{huang2018modeling}, which computes \gls{qoe} as a non-linear function of a combination of coding parameters such as spatial resolution and quantization factor, whose parameters are derived from the spatial activity in the image and the low-order moments of the luminance distribution. 

Learning tools can also be used to estimate these models: in~\cite{yang2017objective}, \gls{bp} is applied on inputs on multiple scales, considering single pixels, regional superpixels, salient objects, and the complete projection, resulting in the \gls{qavr} metric. \glspl{gan} are another learning tool that can be used to train neural networks to estimate \gls{qoe}, and the \gls{deepvriqa}~\cite{kim2019deep} metric is based on them. \glspl{gan} involve two neural networks in opposition to each other: as one network is trained to estimate the \gls{qoe}, the other's objective is to generate examples that trick the other into estimating an incorrect quality. This improves training convergence and can increase overall correlation with subjective test scores. The metric in \cite{li2018bridge} includes head and eye movement data in the learning process, concatenating patch-level \glspl{cnn} with a fully connected network to obtain the \gls{qoe} score. \glspl{cnn} can also be used to determine 3D omnidirectional video quality~\cite{yang20183d}, with additional preprocessing. The \gls{vcnn} model combines viewport prediction with a \gls{cnn}~\cite{li2019viewport}: the \gls{qoe} for different viewports is computed by the \gls{cnn}, while another spherical \gls{cnn} predicts possible future viewports' viewing probability and determine the weights of their contribution to the expected \gls{qoe}.
Table~\ref{tab:qoe} presents a summary of the main full-reference \gls{qoe} metrics presented in this section, along with the references of the comparison studies they appear in.

\begin{table}[t]
	\centering 
	\caption{Summary of the main presented objective \gls{qoe} metrics}
	\label{tab:qoe}
	\setlength{\tabcolsep}{1pt}
	\tiny
	\begin{tabular}{p{2cm}|p{5cm}p{4.5cm}}
		\toprule
		Metric & Description & Comparison studies  \\
		\midrule
		\gls{psnr} & Pixel-level \gls{mse} over the whole image (2D) & \cite{sun2016ahg,alshina2017jvet,zakharchenko2016ahg,tran2017evaluation,kim2019deep,xu2018assessing,tran2017subjective,yang2017objective}
		\\&&\cite{upenik2017on,zhang2011fsim,huang2018modeling,sun2018large,chen2018spherical,croci2019voronoi,zhang2018subjective,duan2018perceptual,tran2018study} \\
		\gls{ssim}~\cite{wang2004image} & Structural similarity on a small scale (2D)  & \cite{zou2017perceptual,xu2018assessing,kim2019deep,zhang2011fsim,huang2018modeling,croci2019voronoi,sun2018large}
		\\&&\cite{yang2017objective,chen2018spherical,zhang2018subjective,duan2018perceptual,tran2018study} \\
		\gls{msssim}~\cite{wang2003multiscale} & Structural similarity on multiple scales (2D) & \cite{upenik2017on,kim2019deep,zhang2011fsim,huang2018modeling,sun2018large,croci2019voronoi,duan2018perceptual,tran2018study} \\ 
		\gls{vifp}~\cite{sheikh2006image} & Shannon model measuring shared information (2D) & \cite{upenik2017on,kim2019deep,huang2018modeling,duan2018perceptual}\\
		\gls{fsim}~\cite{zhang2011fsim} & Feature-based model & \cite{zhang2011fsim,duan2018perceptual}\\
		\midrule
		\gls{spsnr}~\cite{alshina2017jvet} & \gls{psnr} on sampling points from a sphere, remapped on the 2D projection & \cite{alshina2017jvet,xu2018assessing,hanhart2018360,tran2017evaluation,kim2019deep,yang2017objective}
		\\&&\cite{croci2019voronoi,tran2017subjective,upenik2017on,sun2018large,chen2018spherical,zhang2018subjective,li2018bridge,tran2018study}\\
		\gls{wspsnr}~\cite{sun2016ahg} & PSNR weighted proportionally to pixel area on the sphere & \cite{sun2016ahg,xu2018assessing,hanhart2018360,croci2019voronoi,tran2017evaluation,kim2019deep}\\&&\cite{tran2017subjective,upenik2017on,huang2018modeling,sun2018large,chen2018spherical,zhang2018subjective,li2018bridge,tran2018study} \\
		\gls{cpppsnr}~\cite{zakharchenko2016ahg}& Compares quality across projection methods with equal area projection& \cite{zakharchenko2016ahg,hanhart2018360,xu2018assessing,croci2019voronoi,tran2017evaluation}
		\\&&\cite{kim2019deep,tran2017subjective,upenik2017on,sun2018large,chen2018spherical,zhang2018subjective,li2018bridge,tran2018study}\\
		\gls{sssim}~\cite{chen2018spherical} & \gls{ssim} with corrections for projective distortion in the spherical domain & \cite{chen2018spherical}\\
		\gls{wsssim}~\cite{zhou2018weighted} & \gls{ssim} weighted proportionally to pixel area on the sphere &~\cite{zhou2018weighted}\\
		Voronoi~\cite{croci2019voronoi} & \gls{ssim} and \gls{psnr} on Voronoi patches &~\cite{croci2019voronoi}\\
		\gls{cppsnr}~\cite{xu2018assessing} & Saliency- and viewport-weighted \gls{psnr} &~\cite{xu2018assessing}\\
		\gls{cpssim}~\cite{xu2018assessing} & Saliency- and viewport-weighted \gls{ssim} &~\cite{xu2018assessing}\\
		\gls{pvq}~\cite{zou2017perceptual} & Non-linear function of \gls{ssim} & \cite{zou2017perceptual}\\
		\gls{nqq}~\cite{huang2018modeling} & Non-linear function of the coding parameters& \cite{huang2018modeling}\\ \midrule
		\gls{qavr}~\cite{yang2017objective} & Learning-based model based on features at multiple scales & \cite{yang2017objective} \\
		\gls{deepvriqa}~\cite{kim2019deep} & Adversarial generative model to learn \gls{qoe} & \cite{kim2019deep}\\
		Model in~\cite{li2018bridge} & Learning-based metric with head and eye movement input &\cite{li2018bridge}\\
		\gls{vcnn}~\cite{li2019viewport} & \gls{cnn} on viewports weighted by viewing probability & \cite{li2019viewport}\\
		\bottomrule
	\end{tabular}
\end{table}

No reference metrics can measure \gls{qoe} in different context, in which no uncompressed image is available. Metrics such as the \gls{niqe}~\cite{mittal2012making}, based on natural image statistics, and the \gls{sisblim}~\cite{gu2014hybrid}, which is the combination of six different distortion measurements, have good performance on 2D images and videos, but the only study to check their effectiveness for immersive video~\cite{sun2018large} has found that their performance is significantly affected by the geometric distortion, making them only weakly correlated with subjectively perceived quality. The \gls{mc360} metric~\cite{sun2019mc360iqa} is a no reference metric using a multi-channel \gls{cnn} on the six faces of a cube, trained on the dataset in~\cite{sun2018large}: the metric outperforms even 2D full reference metrics on the dataset. 

\subsection{Evaluating QoE metrics}
The conditions for testing \gls{qoe} metrics in immersive video are specified by the \gls{jvet} in~\cite{alshina2017jvet}; a wider discussion on the framework~\cite{hanhart2018360} also provides some reference experiments, with objective and subjective quality metrics; it also introduces the \emph{evil viewport} problem. Evil viewports correspond to \glspl{fov} in which the discontinuous edge caused by the stitching of images from different cameras is clearly visible; it is important to consider evil viewports as a separate case, as \gls{qoe} metrics that take the whole sphere into account might underestimate their impact on \gls{qoe} because of the relatively small area of the stitching edge. Furthermore, another study~\cite{huang2018time} argues that short videos should not be used for \gls{qoe} evaluation in \gls{vr}, as users' attention takes longer to focus in this kind of environment. A detailed evaluation of the \gls{jvet} database, with subjective experiments, is presented in~\cite{zhang2018subjective}.

\begin{table}
	\centering 
	\caption{Performance of the main presented objective \gls{qoe} metrics. The table should be read horizontally: the metric in each row is compared to one for each column. Metrics whose rows have more green cells are more closely correlated with subjective \gls{mos}}
	\label{tab:qoe_vs}
	\tiny
	\aboverulesep=0ex
	\belowrulesep=0ex
	\begin{tabularx}{0.9\textwidth}{|X|X|X|X|X|X|X|X|}
	\toprule
	& \gls{psnr} & \gls{ssim} & \gls{msssim} & \gls{vifp} & \gls{wspsnr} & \gls{spsnr} & \gls{cpppsnr}\\
	\midrule
	\gls{psnr} & \cellcolor{gray!25} & \cellcolor{red!75} Worse & \cellcolor{red!75} Worse & \cellcolor{red!75} Worse& \cellcolor{red!75} Worse & \cellcolor{red!75} Worse & \cellcolor{red!75} Worse \\
	\midrule
	\gls{ssim} & \cellcolor{green!75} Better & \cellcolor{gray!25} & \cellcolor{yellow!50} Similar & \cellcolor{red!75} Worse & \cellcolor{green!75} Better & \cellcolor{green!75} Better& \cellcolor{green!25} Slightly better\\
	\midrule
	\gls{msssim} & \cellcolor{green!75} Better & \cellcolor{yellow!50} Similar & \cellcolor{gray!25} & \cellcolor{red!75} Worse & \cellcolor{green!75} Better & \cellcolor{green!75} Better& \cellcolor{green!25} Slightly better\\
	\midrule
	\gls{vifp} & \cellcolor{green!75} Better & \cellcolor{green!75} Better & \cellcolor{green!75} Better & \cellcolor{gray!25} & \cellcolor{green!75} Better & \cellcolor{green!75} Better & \cellcolor{green!75} Better\\
	\midrule
	\gls{wspsnr} & \cellcolor{green!75} Better & \cellcolor{red!75}Worse & \cellcolor{red!75} Worse & \cellcolor{red!75} Worse & \cellcolor{gray!25} & \cellcolor{red!25} Slightly worse & \cellcolor{red!25} Slightly worse\\
	\midrule
	\gls{spsnr} & \cellcolor{green!75} Better & \cellcolor{red!75} Worse  & \cellcolor{red!75} Worse  & \cellcolor{red!75} Worse  & \cellcolor{green!25} Slightly better & \cellcolor{gray!25} & \cellcolor{red!25} Slightly worse\\
	\midrule
	\gls{cpppsnr} &  \cellcolor{green!75} Better & \cellcolor{red!25} Slightly worse & \cellcolor{red!25} Slightly worse & \cellcolor{red!75} Worse & \cellcolor{green!25} Slightly better & \cellcolor{green!25} Slightly better& \cellcolor{gray!25}\\
	\bottomrule
	\end{tabularx}
\end{table}

In recent years, several studies have compared objective quality metrics to measure their correlation with actual subjective \gls{qoe}: due to the strong dependence of the correlation between objective metrics and \gls{mos} on the actual content of the images, tests performed on different datasets often have contradictory results, and the wide variation across videos of the same dataset confirms that the effect is fundamental and not due to experimental design. The subjective experiments in~\cite{tran2017subjective}, for example, show no advantages of the 360-specific \gls{psnr}-based metrics over the baseline 2D metric; however, this contradicts the results in~\cite{tran2017evaluation, zhang2018subjective}, which both find that \gls{cpppsnr} has better performance than the other metrics, and \gls{spsnr} and \gls{wspsnr} also outperform standard \gls{psnr}. All of the works above~\cite{tran2017evaluation,tran2017subjective} confirm that \gls{mos} decreases sharply if the resolution is lower than 1920p; since only part of the video is inside the viewport at any time, even 1080p video has a low perceived resolution. All later studies confirm that standard \gls{psnr} is worse than any other quality metric, but they often include other metrics, such as \gls{ssim}~\cite{wang2004image} and \gls{vifp}~\cite{sheikh2006image}. In~\cite{upenik2017on,duan2018perceptual,huang2018modeling}, \gls{vifp} significantly outperforms \gls{ssim}, \gls{msssim} and \gls{wspsnr}, which achieve a similar performance, while \gls{psnr} does even worse. Similar results are reported in~\cite{chen2018spherical}, which includes \gls{sssim} but not \gls{vifp} or \gls{msssim}; the 360-specific \gls{ssim} variant outperforms both its 2D ancestor and the \gls{psnr}-based metrics.
The most complete study, which includes several less common 2D \gls{qoe} metrics and \gls{ssim} flavors, finds that \gls{ssim} outperforms both \gls{msssim} and the various \gls{psnr}-based metrics. The results of the various experimental studies are summarized in Table~\ref{tab:qoe_vs}, which compares all the algorithms that are present in at least two of the works presented in this section. The table should be read horizontally: in each row, the corresponding metric is compared to the others (one in each column), and a qualitative summary of the comparison is given by the cell color. The row corresponding to \gls{vifp}, for example, is completely green, showing that it does better than any other metric in the studies in which it is examined, while \gls{psnr}'s row is entirely red. An interesting case is presented by the comparison between \gls{ssim} and \gls{msssim}, whose relative performance is similar, but with a very high variance: \gls{msssim} performs better on some datasets~\cite{upenik2017on}, but worse in others~\cite{sun2018large}, and neither is clearly better in others~\cite{huang2018modeling}. Another work~\cite{tran2018study} compares the basic metrics' performance and complexity, and finds that most are well-correlated to \gls{mos} in the studied scenarios.
The experiments by the authors show that the more complex methods in \cite{zou2017perceptual,huang2018modeling,li2018bridge,yang2017objective,kim2019deep} have a higher performance than traditional metrics, but they have not been corroborated by independent studies yet.

The results of the analyses and comparisons are summarized in Table~\ref{tab:qoe_vs}, with a color-coding scheme to give the reader a first-glance impression of the metrics.

\subsection{Dynamic factors in video QoE}\label{ssec:dynamic_qoe}

The dynamic nature of video is also a major factor in \gls{qoe} that should be taken into account: as in 2D video streaming, stalling events~\cite{kelkkanen2018coefficient} can significantly affect both the perceived quality of 360$^{\circ}$ videos~\cite{kara2016getting} and the sense of presence of the experience~\cite{schatz2017towards}. Since omnidirectional video is more bandwidth-intensive than standard video of the same quality, and buffering is limited by the accuracy of \gls{fov} prediction, as we will discuss in detail in Sec.~\ref{sec:streaming}, avoiding rebuffering events is likely to be a major issue in bitrate adaptation algorithm design.

Quality fluctuations also have an impact on \gls{qoe}, and omnidirectional video can have two sources of picture quality variation: as in all adaptive video streaming systems, the bitrate adaptation algorithm can change the quality to adapt to the connection, either decreasing it if the available bandwidth does not support the current quality level or increasing it if there is unused capacity. The second cause of quality fluctuations is specific to omnidirectional video: as we will discuss in Sec.~\ref{sec:streaming}, streaming systems transmit regions outside the predicted viewport at a lower quality to save bandwidth, which causes sharp decreases in \gls{qoe} when the user turns and the lower-quality content is displayed.

The impact of quality variations due to \gls{fov} changes in adaptive systems is modeled in~\cite{ou2014q}, using quotients between exponential functions of the quality variation rate to approximate the subjective quality when fluctuations are present. This model is extended in~\cite{xie2019modeling}, which considers a more complete model for several different possible scenarios and tests it on a large-scale subjective evaluation dataset. Naturally, a more precise model of the trajectory of the user's gaze could improve the accuracy of these \gls{qoe} models, tying quality evaluation, encoding, and \gls{fov} tracking inextricably.

Another study~\cite{schatz2019tile} investigates the impact of head turn movements on subjective \gls{qoe}, finding that these movements can have a strong impact on perceived quality. However, the effect of user movements on the \gls{qoe} of omnidirectional video is still largely unexplored, and should be investigated further. Another interesting issue, which is explored in~\cite{zhang2018audio}, is the impact of audio degradation on omnidirectional video \gls{qoe}: the authors use a neural network to combine the effects of video and audio impairment, training it on a subjective assessment dataset.

Immersive videos with fast camera motions are also subject to cybersickness~\cite{davis2014systematic}, which is caused by a mismatch between perceived motion and visual input. Cybersickness symptoms often include oculomotor disturbances, nausea, and disorientation, and they are strongly dependent on the content~\cite{kim2019vrsa}: immersive scenarios with strong pitch motion such as rollercoaster rides or parachute dives can induce far stronger symptoms than more horizontal scenes. The technical challenges of designing immersive systems are explored in more detail in~\cite{liu2017360,wanick2018virtual}.

Gaming is another important application of \gls{vr}, and the definition of \gls{qoe} can be slightly different in this context, as both enjoyment and performance need to be taken into account. Immersive gaming is affected both by the quality of the video and by other factors such as the control scheme~\cite{martel2017controlling}, which should include the headset movement input: measurement studies have been performed in different contexts, such as driving simulators~\cite{hupont2015how}, first-person shooters~\cite{lugrin2013immersive}, sport simulators~\cite{wood2017investigating}, or even training simulators~\cite{yue2016evaluation}. 

\section{Saliency and FoV tracking}\label{sec:saliency}

Saliency is the quality that makes part of an image or video stand out and capture viewers' attention~\cite{underwood2006visual}. In this section, we discuss how to evaluate saliency in omnidirectional videos, then apply the concepts to \gls{fov} tracking, which represents not just the importance of parts of images but the trajectory that users' gazes have over the whole duration of the video.

While saliency estimation and \gls{fov} tracking are not, in and of themselves, optimizations that improve the \gls{qoe} of 360$^\circ$ video streaming, they are closely intertwined with all the other components that we discuss in this survey. The most effective projection methods take user behavior into account~\cite{yu2015content}, as prioritizing the content that is watched most often will usually lead to a higher compression efficiency. The same reasoning applies to \gls{qoe} estimation: while we can look at the quality of a 360$^\circ$ frame from all possible angles, the actual experience of users will always entail a single trajectory throughout the video, as their eyes can only look in one direction at a time. Naturally, different users might follow different paths during the videos, looking at different points at different times, and even the same user might focus on different content when rewatching an omnidirectional video, but this makes extensive studies of saliency all the more important.

Finally, \gls{fov} tracking is a key component of streaming systems, as we will discuss in detail in Sec.~\ref{sec:streaming}: since \gls{qoe} only depends on the parts of the video that the user is currently watching, buffer-aided streaming systems can improve their efficiency by predicting which direction the user will look and prefetching the correct parts of the video, or adjusting the projection to improve quality in that direction. A precise, long-term \gls{fov} tracking can then enable the streaming client to make more foresighted choices, 

\subsection{Saliency evaluation}

While there is a wide body of literature on 2D saliency evaluation~\cite{borji2018saliency}, omnidirectional video saliency is still a recent field.
The \gls{bms} and \gls{gvbs} 2D saliency metrics were adapted to omnidirectional images and videos in~\cite{lebreton2018gbvs360}, applying them directly on the omnidirectional images by using the \gls{erp} and automatically compensating for the distortion in the CIELAB color space~\cite{connolly1997study}.
Another attempt to adapt saliency metrics to panoramic video was made in~\cite{startsev2018360}, using similar tools to compensate for the equirectangular distortion. A later work~\cite{sitzmann2018saliency} considers multiple projections, taking into account the bias towards looking at the center of the panorama~\cite{judd2009learning}, i.e., keeping close to the equator of the video sphere~\cite{suzuki2018saliency}, and combining it with 2D metrics. Other saliency metrics, taking center bias and multi-object confusion into account, are proposed in~\cite{ding2018panoramic} and~\cite{nguyen2018your}; the latter also includes a movement tracking framework. A metric considering a linear combination of low-level features and high-level ones such as faces and people was proposed in~\cite{battisti2018feature}, obtaining good results for images containing humans. It is also possible to apply 2D techniques such as weakly supervised \glspl{cnn} directly by using the appropriate projection and adjustments~\cite{cheng2018cube}, or by using \glspl{cnn} to correct the distortion, combining the output of the traditional saliency map of each path with its spherical coordinates~\cite{monroy2018salnet}. Spherical \glspl{cnn} can also be used directly~\cite{zhang2018saliency}.

In~\cite{fang2018novel}, a superpixel decomposition is applied to the image, which is then converted to the CIELAB color space; the difference in contrast and color is then used to train an unsupervised learner to determine saliency, according to the boundary connectivity measure~\cite{yan2018unsupervised}. A similar approach is taken in~\cite{ling2018saliency}, in which the authors derive sparse color features and apply a model of human perception, biased towards the equator, to derive saliency. It is also possible to combine 2D saliency maps on different projections with spherical domain optimization to generate a hybrid metric~\cite{dedhia2019saliency}, or to include illumination normalization \cite{biswas2017towards} to compensate for lighting variations in the omnidirectional images.
\glspl{gan}~\cite{chao2018salgan} are another supervised learning tool that can be used to infer saliency; unsupervised learning from bottom-up features has also been applied successfully~\cite{xia2016bottom}. An experimental comparison of several standard and omnidirectional state-of-the-art saliency detection techniques is presented in~\cite{ozcinar2018visual}.

Scanpaths~\cite{cerf2007predicting} are a natural extension of the saliency metric, adding the time dimension to the static map; image metrics can often be straightforwardly extended to the video domain, both for standard and omnidirectional video~\cite{lebreton2018v}. Scanpaths can also act as predictors of future gaze directions when used as the training model for learning agents such as deep networks~\cite{assens2018scanpath} or \glspl{gan}~\cite{assens2019pathgan}. However, scanpath models often have the same issues as static saliency models: since saliency is extremely content-dependent, different models can have higher performance on different datasets. For this reason, standard evaluation datasets and metrics have been proposed~\cite{gutierrez2018introducing,gutierrez2018toolbox}.
In~\cite{xie2018cls}, an approximate saliency metric is derived by clustering multiple users' head movements, but the training is video-specific and does not generalize on other content. A more general model based on user movement statistics is derived in~\cite{de2017look} by combining \glspl{fsm}~\cite{chang2011fusing} with head movement data and applying an equator bias.

In general, saliency evaluation is more related to coding and compression than to streaming, as streaming systems have the benefit of knowing the current trajectory of the user, which can lead to more effective \gls{fov} tracking tools discussed below. On the other hand, the compression and coding phase must be performed once, so saliency and most frequent scanpath estimation are the only available tools to use content information during it. As with other fields, the development of machine learning tools to combine content features and user experience is one of the major research challenges: the field is rapidly developing, and a one-step network that can automatically learn to extract saliency and encode the video at the same time is just behind the corner.

\begin{table}
	\centering 
	\caption{Summary of the main presented saliency \gls{fov} prediction methods}
	\label{tab:fov}
	\tiny
	\begin{tabular}{p{1cm}|p{4.5cm}p{4.5cm}} 
		\toprule
		Reference & Type & Basic principle  \\
		\midrule
		\cite{assens2019pathgan} & Content- and popularity-based & \gls{gan}\\
		\cite{xie2018cls} & Popularity-based & Clustering\\
    \cite{ramanathan2007rate} & History-based & Dead reckoning\\
    \cite{singhal1995exploiting} & History-based & Polynomial regression\\
    \cite{kiruluta1997predictive,aykut2018delay} & History-based & Kalman filtering\\
    \cite{bogdanova2010dynamic,feng2019viewport} & History- and popularity-based & Gaussian filtering\\
    \cite{petrangeli2018trajectory} & History- and popularity-based & Clustering\\
    \cite{zou2019probabilistic} & History- and popularity-based & \gls{cnn}\\
    \cite{fan2019optimizing} & History- and popularity-based & \gls{rnn}\\    
    \cite{fan2019optimizing,fan2017fixation,li2018two, xu2018gaze} & Content-, history- and popularity-based &\gls{lstm}\\
    \cite{li2019very} & Content-, history- and popularity-based & Convolutional \gls{lstm}\\
    \cite{yu2019field} & Content- and history-based & Attention-based encoder-decoder network\\
		\bottomrule
	\end{tabular}
\end{table}

A task related to saliency and scanpath estimation is automatic navigation, i.e., moving through a panoramic video to catch the most important parts of the action.
A simple optimization is performed in~\cite{maugey2017saliency}, while another work~\cite{hu2017deep} proposes a combination of object recognition and reinforcement learning, implementing the policy gradient technique to track interesting objects in sports videos. A similar approach can be applied to explore a space by rewarding an agent when it examines unexplored portions of its environment~\cite{jayaraman2018learning}.

\subsection{Field of View prediction}\label{ssec:prediction}

As discussed in Sec.~\ref{sec:quality}, the viewport direction is a fundamental factor in assessing the \gls{qoe} of immersive video, and needs to be considered proactively both in the coding phase and when performing adaptive streaming. In particular, the difficulty of predicting future viewport orientation leads to diminishing returns on capacity, limiting the amount of prefetching~\cite{almquist2018prefetch} and exposing users to the risk of annoying stalling events~\cite{schatz2017towards}. 

The prediction of gaze direction has been studied since the '90s by using simple analytical tools, and it parallels the work on motion prediction: the first studies used dead reckoning~\cite{ramanathan2007rate} and polynomial regression~\cite{singhal1995exploiting}, and several streaming systems that exploit \gls{fov} prediction still apply simple linear regression on historic data~\cite{bao2016shooting}. However, the models are often too simplistic, not capturing viewer behavior complexity: an early frequency-domain analysis~\cite{azuma1995frequency} highlights the difficulty of predicting long-term trends using these strategies. Kalman filtering approaches use similar underlying models, but they can deal with imprecise measurements of the orientation~\cite{kiruluta1997predictive,aykut2018delay}.

Recently, more complex statistical tools such as Gaussian filtering~\cite{bogdanova2010dynamic,feng2019viewport} and clustering~\cite{petrangeli2018trajectory} have been used with good results, modeling viewer gaze direction as a random variable whose distribution is determined by their own history as well as past users' behavior.
Another study on the correlation in the behavior of users~\cite{carlsson2019had} concentrates on the caching implications of predicting \gls{fov}.

Recently, deep learning has also been applied to the problem, as \gls{fov} prediction is a classical regression problem: both \glspl{cnn}~\cite{zou2019probabilistic}, and \glspl{rnn}~\cite{fan2019optimizing} had good performance on standard datasets~\cite{upenik2017simple}. Three other works~\cite{fan2017fixation,li2018two, xu2018gaze} introduce \glspl{lstm}, including content-related metrics such as saliency maps and scanpaths along with the motion information. In~\cite{zhao2019laddernet}, ladder convolution is used before the \gls{lstm} to extract contextual information from the encoded image and correct for the projection. Naturally, a richer state with more information from different sources can improve the quality of the prediction, which is further enhanced in~\cite{yu2019field} by the use of an encoder-decoder network with an attention mechanism that can have high tracking accuracy over multiple seconds. However, these methods have not been tested on large datasets yet, and their significant computational complexity poses a challenge in real-time mobile applications. The search for an efficient \gls{fov} tracking algorithm that can allow \gls{dash} clients to achieve similar levels of buffer filling to traditional planar video is still open, and as these works are all from the past 3 years, the state of the field is rapidly changing and improving.

Prediction on even longer timescales is possible by leveraging the watching history of other users and identifying similarities~\cite{li2019very}, maintaining a viewport hit rate over 75\% even at a distance of 10 seconds. For additional accuracy, users can be clustered by similarity~\cite{ban2018cub360,rossi2019spherical}, identifying common patterns within clusters more effectively. This approach can also be combined with deep reinforcement learning~\cite{xu2018predicting} to reduce training costs. It is also possible to use combine saliency metrics and head movement with more precise gaze tracking, obtaining a higher precision in the prediction~\cite{zhu2018prediction}. 
\gls{fov} prediction can also be tested on public datasets, often used by existing saliency estimation~\cite{ozcinar2018visual} and prediction methods~\cite{xu2018predicting}; the latter provides a dataset with the head movements of 58 users across 76 video sequences. The datasets used for \gls{qoe} measurement often include both the ratings and head movements of the viewers, so they can also be used for this purpose. A dataset with the head movements of 59 users watching 7 YouTube immersive videos was presented in~\cite{corbillon2017360}, while another dataset with partly overlapping videos and 50 different subjects was presented in~\cite{lo2017360}. Another dataset includes the head trajectories of 48 users watching 18 videos~\cite{wu2017dataset}, and yet another~\cite{nguyen2019saliency} contains the \gls{fov} trajectories and saliency maps of 48 users on 24 videos. The dataset presented in~\cite{fremerey2018avtrack360} includes both head movements and the results of a cybersickness questionnaire for 20 subjects watching 48 video sequences. The same kinds of data are available in~\cite{nasrabadi2019taxonomy}, with 60 subjects watching 28 videos, and in~\cite{knorr2018director}, with 20 subjects watching 5 videos created and edited by professional filmmakers. Another dataset~\cite{rai2017dataset} provides eye tracking data, which is more precise than head movements, for 98 static images, observed by 63 subjects for 25 seconds each. Viewer gaze direction is usually analyzed on \gls{vr} headsets, but there is a public dataset~\cite{duanmu2018subjective} of immersive video \glspl{fov} on a desktop platform. The datasets on \gls{fov} prediction and tracking are summarized in Table~\ref{tab:fov_datasets}, while the main methods of \gls{fov} prediction we presented in this section are summarized in Table~\ref{tab:fov}.

\begin{table}
	\centering 
	\caption{Available \gls{fov} tracking datasets}
	\label{tab:fov_datasets}
	\tiny
	\begin{tabular}{llll}
		\toprule
		Reference & Type & Subjects & Videos\\
		\midrule
    \cite{xu2018predicting} & Head movements & 58 & 76\\
    \cite{corbillon2017360} & Head movements & 59 & 7\\
    \cite{lo2017360} & Head movements & 50 & 10\\
    \cite{wu2017dataset} & Head movements & 48 & 18\\
    \cite{nguyen2019saliency} & Head movements (with saliency maps) & 48 & 24\\
    \cite{fremerey2018avtrack360} & Head movements (with cybersickness questionnaire) & 20 & 48\\
    \cite{nasrabadi2019taxonomy} & Head movements (with cybersickness questionnaire) & 60 & 28\\
    \cite{knorr2018director} & Head movements (with cybersickness questionnaire) & 20 & 5\\
    \cite{rai2017dataset} & Eye movements (static images) & 63 & 98\\
    \cite{duanmu2018subjective} & Eye movements (desktop platform) & 50 & 12\\
		\bottomrule
	\end{tabular}
\end{table}

\section{Streaming}\label{sec:streaming}

Serving omnidirectional video content over the Internet is a complex problem of its own: a naive approach sending the whole sphere at the highest quality will be extremely inefficient, and an intelligent way to adapt to network conditions and user behavior needs to be devised. In this section, we discuss the standardization work on omnidirectional video streaming and the solutions to optimize bitrate adaptation by considering spatiotemporal elements such as \gls{fov} prediction. Finally, we present some of the work on network support of omnidirectional video in the context of \gls{vr}, which is one of the key applications that will be enabled by 5G networks. 

\subsection{Streaming standardization}\label{ssec:standard}
Today, the \gls{dash} streaming standard is almost universally used for 2D video streaming over the Internet: it divides videos into short segments, which are encoded independently and at several different qualities by the server. The streaming client can then choose the quality level for each segment, depending on the bitrate its connection can support, by requesting the appropriate HTTP resource. The low computational load on the server and transparency to middleboxes make \gls{dash} highly compatible with the existing Internet infrastructure, and the possibility of implementing different adaptation algorithms makes it versatile to different network conditions. In the early 2010s, the standard was extended to enable the transmission of omnidirectional, zoomable and 3D content: the \gls{srd} extension~\cite{niamut2016mpeg} specifies spatial information on each segment, allowing servers to present spatially diverse content. The standard only specifies the spatiotemporal coordinates of each segment, and the choice of which ones to download and show to the user is still client-side, in accordance with the client-based \gls{dash} paradigm.

The \gls{omaf} standard~\cite{hannuksela2019overview} is another specification that can extend \gls{dash} or other streaming systems by specifying the spatial nature of video segments. Furthermore, \gls{omaf} also specifies some requirements for players, taking another step towards a complete standard specification for omnidirectional streaming. In fact, \gls{omaf}-based players have already been implemented and demonstrated~\cite{skupin2017viewport}. The standard specifies a viewport-independent video profile using the \gls{hevc} coding standard, as well as two viewport-dependent profiles using \gls{hevc} or the older \gls{avc}, supporting the \gls{erp} and \gls{cmp} projections and tile-based streaming. \gls{omaf} further defines a viewport-dependent projection approach, in which the client chooses the projection with the highest quality for its current viewport, as well as three different tile-based streaming approaches: in the simplest one, the viewport region is downloaded at a high quality, along with an additional low-quality version of the whole sphere. The other two allow a freer choice by the client, which can download a set of tiles with either mixed encoding quality or mixed resolutions, privileging the viewport area in both cases.

A \gls{dash} \gls{srd} or \gls{omaf} compliant server can allow clients to stream omnidirectional video, presenting either segments with different viewport-dependent projections or separate tiles for the client to choose. The client can download the appropriate projected content, potentially discarding or downloading low-quality versions of tiles with a low viewing probability and saving bandwidth. It is also possible to exploit the features of \gls{hevc} to enable fast \gls{fov} switching or to give users the option to zoom into certain areas of the sphere~\cite{d2016using}, as high-quality chunks can be requested at any moment if the user moves their head~\cite{song2019fast}, seamlessly integrating the functions with minimal server-side changes. The techniques for streaming content at the highest possible quality exploiting viewport information are described in detail in the following.

\subsection{Viewport-dependent streaming}\label{sec:viewport_stream}

Omnidirectional streaming has all the complexity of traditional streaming, with buffer concerns and dynamic quality considerations, but it has an additional degree of freedom: since the viewer only sees the portion of the sphere in their \gls{fov}, quality is strongly dependent on the direction of their gaze~\cite{nguyen2017impact}. the parts of the sphere inside the \gls{fov} are visualized by the user, and their attention focuses on a narrower foveal cone~\cite{lungaro2017gaze}. Naturally, adaptive streaming systems try to exploit this by maximizing the quality of the predicted \gls{fov} at the expense of unwatched regions, which do not contribute to the \gls{qoe}. This approach is not without pitfalls: standard \gls{dash} buffered streaming often prefetches segments several seconds in advance, with no performance loss, but prefetching an unwatched region at a high quality does not lead to any \gls{qoe} improvement~\cite{he2018joint}, so the advantages of prefetching in adaptive 360$^\circ$ video are closely tied with the quality of the viewport prediction~\cite{almquist2018prefetch}.
The paradigm can also deal reactively to dynamic viewpoint changes~\cite{corbillon2018dynamic}. 

Transmission factors can significantly affect the quality of the image~\cite{schatz2017towards}: viewport-agnostic streaming, which transmits the whole omnidirectional video with the same quality, does not introduce additional distortion, but it is extremely bandwidth-inefficient. There are two viewport-dependent approaches to adapting omnidirectional streaming systems to the \gls{fov}. The first, and most common, approach is tile-based streaming, which divides the omnidirectional video into independent rectangular tiles~\cite{hosseini2016adaptive}. In this case, the bitrate adaptation becomes multi-dimensional~\cite{petrangeli2017http}: each tile can be streamed independently at a different quality level, and the client reconstructs the whole sequence. It is also possible to exploit the HTTP/2 weight parameter to control the tile interleaving and prioritization~\cite{yahia2018http}. The main downsides of tiling-based approaches are the frequent spatial quality fluctuations~\cite{concolato2017adaptive} and artifacts close to tile borders. 

The second approach is viewport-dependent projection, which uses offset projection~\cite{sreedhar2016viewport} or differentiated \gls{qp} assignment~\cite{de2019delay} to improve the quality of the \gls{fov}~\cite{corbillon2017optimal}. This approach avoids obvious seams between tiles at different qualities. However, it can have temporal quality fluctuations as the projection changes when the user moves their head, and it is rarely used in the literature because of the server-side memory requirements of storing several different projections with different encoding parameters. A third, even less common, solution in wireless channels is to transmit the video directly, using analog modulation after applying the \gls{dct}~\cite{fuiihashi2018graceful}. This leads to a more graceful quality decrease than the sharp fall caused by digital transmission, but is not without its disadvantages, as the transmitter and receiver hardware need to be designed \emph{ad hoc}. 

In the following, we concentrate on tile-based streaming methods, as they are by far the most common, although they involve a higher computational costs due to the necessity of stitching~\cite{gutierrez2018toolbox}. While the simplicity in the design of tile-based systems is attractive, we remark that they might not be optimal in terms of encoding efficiency, and a more holistic solution that takes both encoding efficiency and streaming factors into account might provide an even better solution in the future. As we discussed in the previous sections, the design of projection and encoding methods is inextricably linked to the expected scanpath of the user's gaze, while the streaming adaptation strategy strongly relies on \gls{fov} prediction. As some users might behave in an atypical manner and follow uncommon scanpaths, the encoding system and streaming systems need to guarantee a minimum \gls{qoe} in all cases, while optimizing the \gls{qoe} for as many users as possible. These conflicting objectives present an interesting trade-off, which is mostly unexplored in the current literature and would be extremely interesting to investigate.

An accurate prediction of the \gls{fov} can improve the efficiency of omnidirectional streaming significantly: since the only area that the viewer sees is the one in the viewport, other parts of the video sphere can be streamed with a much higher compression, or even discarded, without affecting the \gls{qoe}. Several authors have proposed streaming algorithms exploiting this prediction, often using it in one of two ways:
\begin{itemize}
 \item The \emph{viewport-based} approach maximizes the quality of the predicted \gls{fov}, or a slightly wider region to account for inaccuracies in the prediction, and streaming the rest of the sphere at the lowest quality.
 \item The \emph{probabilistic} approach weights the tiles by their viewing probability, then optimizing the expected quality.
 \item The \emph{reinforcement learning} approach implicitly optimizes the expected long-term \gls{qoe} by applying its namesake learning paradigm.
\end{itemize}
Naturally, the capacity of the connection is the constraint that limits the \gls{qoe}, and various capacity prediction methods can be employed. Since there is no correlation between the capacity of the channel and the viewport orientation, the two predictions can be performed separately with different methods, and the use that the streaming adaptation algorithm makes of the results is usually not constrained by the prediction method. An interesting way to improve the prediction and the streaming quality is to devise the content in a way that implicitly or explicitly leads users to direct their attention in certain directions~\cite{sassatelli2019new}.

The viewport-based approach is simpler, as it does not require solving a complex optimization problem: there are only two regions, the one around the viewport and the rest of the sphere, and the second one is usually either not streamed at all or streamed at the maximum possible compression~\cite{he2018rubiks}. Naturally, the approach is optimal if the predictor is perfect. In \cite{bao2016shooting}, both a linear regression and a neural network-based prediction are tested with a simple algorithm that transmits a circular portion of the omnidirectional video, comprised of the circle inscribing the predicted viewport with an additional safety margin. The authors assume that an efficient projection method is used and that capacity is constant. It is also possible to adapt the safety margin to the estimated prediction error variance~\cite{nguyen2017new}, increasing the area in case of quick head movements or highly unreliable predictions. Naturally, linear regression is not the only possible model: a second-degree model with constant acceleration is proposed in~\cite{nguyen2018predictive}, and \gls{svr} with eye tracking data is used in~\cite{yang2019fovr}. The latter distinguishes a small attention area of about 10$^\circ$ close to the gaze direction, while the rest of the \gls{fov} is a larger sub-attention area. The two areas have different weights in the optimization, and a third area (non-attention) completes the sphere with the unwatched portions. This kind of three-tier optimization is a first step towards the probabilistic approach.

It is also possible to mix a popularity-based approach with linear regression: the scheme presented in~\cite{qian2016optimizing} uses the two at the same time, weighting the regression outputs by the popularity and fetching the predicted viewport tiles, with some margin for errors, at the highest quality supported by the connection. A more refined server-side approach is adopted in~\cite{bao2017motion}, which uses a neural network to estimate the future viewport of multiple users. The algorithm then sends the data for the predicted viewport to each user at the highest possible quality, while sending the invisible parts of the sphere at the lowest one to save bandwidth. Another work ~\cite{zou2019probabilistic} takes the same approach, replacing the fully connected neural network with a \gls{cnn}. Object tracking is another kind of information that can be used for the prediction: this semantic information~\cite{leng2018semantic} is often correlated to users' viewing patterns, as their gaze follows one of the object across the panoramic video.

The probabilistic streaming approach weights the quality of each tile by their viewing probability and optimize expected quality assuming constant capacity. This scheme has been combined with linear and ridge regression for the equirectangular~\cite{nguyen2019optimal}, triangular~\cite{qian2018flare}, and truncated pyramid~\cite{xu2018probabilistic} tiling schemes. In all three cases, the capacity of the connection is assumed to be constant. In~\cite{xie2017360probdash}, the linear regression is combined with a buffer-based streaming approach to maintain playback smoothness, adapting the estimate of the total bitrate to control the buffer level. Bas-360~\cite{xiao2018bas} is another scheme which combines spatial adaptation with a temporal factor, optimizing a sequence of multiple future frames together and using stream prioritization and termination to correct bandwidth and \gls{fov} prediction errors. A similar method~\cite{ban2017optimal} considers both temporal and spatial quality smoothness in the optimization, considering a sequence of future segments. The \gls{opv} scheme~\cite{lin2019opv} tackles prediction error from a different angle, correcting its decisions by streaming higher-quality tiles for already buffered segments if necessary. This allows the client to keep a long buffer and avoid stalling without having to lower quality.

As for the viewport-based approach, popularity can be considered to perform the prediction: a proposed scheme~\cite{chakareski2018viewport} tries to maximize the overall expected~\gls{qoe}, considering only the popularity of each tile, corrected for the equirectangular tiling (if the viewport is closer to the poles, more tiles will be part of the \gls{fov}). The algorithm considers the rate-distortion curve for each tile, weighted by its corrected navigation probability. In this case, capacity is assumed to be constant. This approach can also exploit the popularity of tiles and linear regression jointly: in~\cite{koch2019transitions}, a transition threshold between the two methods is set, and the popularity-based model is used if the measured capacity of the connection is insufficient to support the other one. The concept behind this scheme is that regression incurs a higher risk of rebuffering events in low-bandwidth scenarios, and switching to a more conservative scheme is desirable in this context. Another work~\cite{rossi2017navigation} mixing the two prediction methods uses a linear combination of the two outputs, considering the trade-off between the flexibility of the adaptation and the coding efficiency, which decreases as the number of tiles grows. A \gls{knn} was exploited in~\cite{xu2018tile} to make use of previous users' data by finding similar scanpaths and assigning future \glspl{fov} from those users a larger probability.

\begin{table}[t]
	\centering 
	\caption{Summary of the main presented \gls{fov} prediction-based streaming schemes}
	\label{tab:streaming_fov}
	\setlength{\tabcolsep}{4pt}
	\tiny
	\begin{tabular}{p{1cm}p{1.5cm}|p{4cm}p{4cm}} 
		\toprule
		Ref. & Projection & Optimization & Prediction method  \\
		\midrule
		\cite{bao2016shooting} & Ideal & Circular region around the viewport & Linear regression and neural networks \\ 
		\cite{nguyen2017new} & \gls{erp} & Adaptable region around the viewport & Linear regression \\ 
		\cite{nguyen2018predictive} & \gls{cmp} & Highest quality for predicted viewport & Second-degree regression \\ 
		\cite{yang2019fovr} & \gls{erp} & Attention-based weights  & \gls{svr} with eye tracking \\ 
		\cite{qian2016optimizing} & \gls{erp} & Highest quality for predicted viewport & Popularity-weighted linear regression\\ 
		\cite{bao2017motion} & \gls{erp} & Highest quality for predicted viewport & Neural network with motion history\\ 
		\cite{zou2019probabilistic} & \gls{erp} & Highest quality for predicted viewport & \gls{cnn} with motion history\\ 
		\cite{leng2018semantic} & Direct & Highest quality for predicted viewport & Semantic object tracking\\
		\midrule
		\cite{nguyen2019optimal} & \gls{erp} & Expected quality & Linear regression  \\
		\cite{qian2018flare} & Triangular & Expected quality & Linear and ridge regression  \\
		\cite{xu2018probabilistic} & \gls{tsp} & Expected quality & Linear regression  \\ 
		\cite{xie2017360probdash} & \gls{erp} & Expected quality with buffer control& Linear regression\\
		\cite{xiao2018bas} & \gls{erp} & Expected quality over multiple future steps& Unspecified\\
		\cite{ban2017optimal} & \gls{erp} & Expected quality over multiple future steps& Unspecified\\
		\cite{lin2019opv} & \gls{erp} & Expected quality, past action fixes& Unspecified\\
		\cite{chakareski2018viewport} & \gls{erp} & Expected quality & Popularity-based model\\
		\cite{koch2019transitions} & \gls{erp} & Expected quality & Popularity/linear regression switching\\
		\cite{rossi2017navigation} & \gls{erp} & Expected quality & Popularity/linear regression linear combination\\		
		\cite{xu2018tile} & \gls{sp} & Expected quality & \gls{knn} with other users' patterns\\
		\cite{fan2017fixation} & \gls{erp} & Expected quality & \gls{lstm} with saliency, motion, and \gls{fov} info\\ 
		\cite{park2019advancing} & \gls{erp} & Expected quality & 3D-\gls{cnn} with saliency, motion, and \gls{fov} info\\
		\cite{ghosh2018robust} & \gls{erp} & Minimum visible quality, stalling avoidance & Unspecified\\  
		\midrule
		\cite{fu2019360srl} & Unspecified & Reinforcement learning & Unspecified\\ 
		\cite{kan2019deep} & \gls{erp} & Reinforcement learning & Neural network from~\cite{ozcinar2019visual}\\ 
		\cite{jiang2018plato} & \gls{erp} & Reinforcement learning & \gls{lstm}\\ 
		\cite{xiao2019deep} & \gls{erp} & Reinforcement learning & \gls{lstm}\\ 
		\cite{zhang2019drl360} & \gls{erp} & Reinforcement learning & Implicit in the solution\\ 
		\midrule
		\cite{xiao2017optile,nguyen2019client} & Adap. \gls{erp} & Expected quality & Known \gls{fov}\\
		\cite{ozcinar2019visual} & Adap. \gls{erp} & Expected quality & Popularity-based model\\		
		\cite{dunn2017resolution} & Adap. & Expected quality & Popularity-based model\\		
		\bottomrule
	\end{tabular}
\end{table}

A more sophisticated approach, presented in~\cite{fan2017fixation}, combines saliency and motion information with the \gls{fov} scanpath using an \gls{lstm}. The predicted viewing probability for each equirectangular tile can then be used in the usual probability-weighted quality optimization. The same technique was compared to a 3D-\gls{cnn} approach in~\cite{park2019advancing}: both prediction methods had extremely good performance, but the latter had a slight advantage. 

A complete streaming algorithm, which considers stalling and a more sophisticated capacity prediction method based on the harmonic mean of past samples, is presented in~\cite{ghosh2018robust}. The authors derive an efficient heuristic that can maintain a high quality even when the \gls{fov} is uncertain, optimizing the quality of the worst tile in the viewport to guarantee a minimum \gls{qoe} while limiting stalling. However, they do not present a specific \gls{fov} prediction method, but analyze performance as a function of the prediction error.

The third way to achieve the same objective without explicitly optimizing the expected \gls{qoe} is to use \gls{drl}: the sequential approach reduces the multi-dimensional tile quality decision to a sequence of decisions for each single tile~\cite{fu2019360srl}. Another \gls{drl} solution~\cite{kan2019deep} models the problem as a \gls{mdp}, optimizing a complex function considering the \gls{fov} picture quality, quality variations, and stalling events. The work assumes that \gls{fov} prediction is performed by a neural network, as in~\cite{bao2017motion}, and includes the prediction in the model state, along with the capacity and buffer history. Plato~\cite{jiang2018plato} is another system that assumes an external prediction as input to a \gls{drl} system, in this case performed by an \gls{lstm}. A similar solution was presented in~\cite{xiao2019deep}, modeling buffer overflows explicitly. Another work using \gls{drl}~\cite{zhang2019drl360} performs the \gls{fov} prediction implicitly, using an \gls{lstm} to keep track of the historical trends in capacity and viewport orientation.

It is also possible to adaptively change the projection: in~\cite{xiao2017optile,nguyen2019client}, the compression or size of the tiles of an \gls{erp} can be changed according to the user's expected behavior and the expected quality resulting from each scheme. While the authors assume that the future \gls{fov} is known in advance, which is obviously unrealistic, this kind of scheme adds a degree of freedom to the streaming optimization. It is also possible to use the adaptive projection with popularity-based prediction, as in~\cite{ozcinar2019visual}. In~\cite{dunn2017resolution}, the popularity-based prediction is used to derive an adaptive projection with an irregular shape. The trade-off between changing the compression of the tiles at the same resolution and lowering the resolution to increase the bandwidth efficiency has also been explored~\cite{curcio2017bandwidth}, and the results show that the viewport-based approach has a higher \gls{qoe} with the same compression.

Techniques based on packet-level coding or \gls{svc}~\cite{duanmu2017view,nasrabadi2017adaptive} are also possible: a scheme that protects immersive video data with fountain codes, increasing the redundancy for areas in the \gls{fov} while leaving unwatched areas of the sphere unprotected, has been proposed in~\cite{lv2018unequal}.
In a multipath wireless scenario in which multiple links with fast-varying capacity are available, it is possible to use a wireless path to transmit the video's base layer and another to to transmit enhancement layers, improving the quality of live \gls{vr} streaming while maintaining full reliability~\cite{sun2018multi}. 

\subsection{Network-level innovations}\label{ssec:network}

The \gls{dash} paradigm is entirely end-to-end, and does not require any network support. However, several studies have explored the possibility of implementing explicit network support for video streaming: the network can either explicitly communicate with the client and help it make decisions, or provision resources and indirectly improve the situation perceived by the client, which will then improve the video quality autonomously. Since immersive streaming requires more resources from the network, implicit or explicit support is even more helpful in this scenario.

The most basic form of network support for immersive video is at the design level: the lower layer protocols and their interplay can negatively affect the 360$^{\circ}$ stream, and design adjustments based on an analysis of these effects can significantly improve performance. Such a study was performed for the LTE network~\cite{tan2018supporting}, finding several simple solutions that can be implemented without changing the network architecture. The standardization of the 5G requirements and solutions for immersive and \gls{vr} video streaming are ongoing~\cite{gabin20185g}.

Caching is another form of basic network support that can be implemented simply, and is often already in place thanks to \glspl{cdn}. Explicitly considering the nature of immersive video can significantly enhance the efficiency of edge caching strategies~\cite{mahzari2018fov,maniotis2019tile}: by caching the most common fields of view closest to the network edge~\cite{liu2019joint}, it is possible to increase the cache hit rate and, consequently, the average \gls{qoe}. Caching can be combined with edge computing strategies to improve the \gls{qoe} of \gls{ar}~\cite{chakareski2017vr}, rendering the virtual content in the user's \gls{fov} without the latency that cloud processing entails. It is also possible to extend these techniques, along with a measure of user popularity at any given moment, to optimize multicast immersive streaming in mobile networks~\cite{ahmadi2017adaptive}.

More explicit approaches aim at resource allocation when multiple \glspl{rat} are available~\cite{huang2019utility}, exploiting \gls{fov} prediction to pair users with access points and effectively use wireless resources. The same optimization can be performed for multiple users on the same network, maximizing the overall \gls{qoe} by cooperatively downloading different \gls{svc} layers~\cite{zhang2018cooperative}. \gls{fov} prediction can also be used in multicast scenarios, clustering users with similar points of view and exploiting mmWave multicast~\cite{perfecto2018taming} to serve them together. With the gradual adoption of 5G technology, it is also possible to combine cellular resource scheduling optimization with encoding tile rate selection~\cite{yang2019content} to provide low delay upload of \gls{vr} content.

Live streaming of \gls{ar} and \gls{vr} content is another issue, which is complicated by the limited delay tolerance: experimental studies~\cite{grzelka2019impact, mania2004perceptual} show that any delay over 10 ms can be perceived by users as annoying, although higher latencies can be tolerated~\cite{albert2017latency}. The issue becomes even more complex when viewport-adaptive schemes are taken into account, as the adaptation scheme needs to react fast enough to changes in the \gls{fov} to avoid quality drops~\cite{de2019delay}. Future networks need to be able to guarantee reliable end-to-end communication below this latency, requiring innovation both from the physical~\cite{elbamby2018toward} to the transport layer~\cite{chiariotti2019analysis} to enable these applications.

However, network support is not limited to communication: in the case of rendered \gls{vr}, the network can also help with computation tasks. Most \gls{vr} platforms are tethered, using a desktop computer to render the environment in real-time: current smartphones do not have the computing and battery power to provide a high-quality \gls{vr} experience without offloading some of the computational load~\cite{lo2018edge}. Several works have tried to mitigate the latency problems caused by the remote rendering, either by reducing the throughput using compression~\cite{liu2018cutting} or by using servers close to the network edge~\cite{shi2019mobile}. The Furion platform~\cite{lai2019furion} tries to solve this issue by using \gls{fov} prediction techniques to prefetch rendered background content from a remote server, rendering only the foreground objects locally. The use of \gls{mec} to provide rendering support to multiple \gls{vr} users at the same time has also been investigated~\cite{li2018muvr}. The several components of latency in a \gls{vr} application were analyzed in~\cite{mangiante2017vr}: the trade-off between network and computation delay, as cloud servers are more powerful but farther away, is a critical design choice for future systems.

\section{Conclusions and open challenges}\label{sec:concl}

Omnidirectional video has gained significant traction, both in the research community and in the industry, and the first commercial \glspl{hmd} are now several years old. This kind of video presents challenges that call for a redesign of the whole video coding, streaming and evaluation pipeline, taking into account two critical aspects specific to 360$^\circ$ video: geometric distortion due to the mapping of a spherical surface to 2D planes, and the fact that viewers only experience a limited \gls{fov}.

In this survey, we analyzed all aspects of omnidirectional video coding and streaming. First, we reviewed projection methods and the geometric distortion that they can cause, with a description of their effects on video encoders and their compression efficiency. The choice of a projection scheme is often a trade-off between different types of distortion: while approaches based on solids with a larger number of faces approximate the spherical nature of the image better, they also increase the number of edge distortion, and thus the possibility of visible errors at the seams. The same is true for offset projection: dedicating more pixels to the most probable view increases the average \gls{qoe}, but highly reduces it if the user turns around unexpectedly. The subsequent encoding parameters also have effects on the image quality, and they should be optimized jointly with the projection settings.

The projection and encoding of omnidirectional videos is a critical procedure, as it determines the rate-distortion efficiency of the video streaming system. The research on the subject has evolved far from the first simple examples using simple projection schemes and the 2D encoding pipeline, but some fundamental trade-offs limit possible performance. In particular, the choice of projection affects the rest of the encoding pipeline significantly, and \emph{ad hoc} region-adaptive quantization schemes need to be devised. Motion models and inter-frame compression also need to be carefully tuned, as no projection can avoid geometric distortion and discontinuities caused by objects crossing face boundaries at the same time. 

We then focused on \gls{qoe} in omnidirectional video: as several subjective studies prove, 2D quality metrics are inaccurate in this scenario, and more intelligent ones that take geometric distortion and viewer attention are needed. The dynamic factor also plays a role, as quality variations between segments and tiles can affect \gls{qoe} in unpredictable ways. In general, measuring \gls{qoe} in omnidirectional video is a complex problem, and will probably require the use of content-aware learning tools. We then discussed automatic saliency estimation and \gls{fov} prediction techniques, which have a critical role in \gls{qoe} estimation and video streaming: being able to predict the \gls{fov}, both for the average user and for the current viewing session, can help compress video better by allocating more pixels to regions with more important content and which are viewed more often, but also increase the efficiency of tile-dependent streaming and the accuracy of \gls{qoe} metrics. 

The strong dependence between video content and the effectiveness of different metrics, along with the lack of a single large-scale database of experimental results to use, can result in contradictory evidence, and multiple studies often have different outcomes. However, there are a few guidelines for future research: the inadequacy of 2D metrics such as \gls{psnr} in the omnidirectional video domain is evident from most studies, even when corrected and weighted to account for the different geometry. \gls{vifp} seems to be a promising base to develop better omnidirectional \gls{qoe} metrics, but the hot topic in the field is machine learning: a few learning-based metrics have already been proposed, but they have not been tested on a wider scale or released publicly. Whether the significant performance improvements that machine learning achieved in other applications can be replicated in \gls{qoe} measurement of omnidirectional video is arguably the biggest open question. Another important, and often overlooked, factor is the dynamic nature of video, which can be crucial in omnidirectional video due to the cybersickness issue: the study of dynamic metrics for omnidirectional video taking stalling events and quality fluctuations due to the adaptive streaming and the user's head movements into account is still limited to a few works.

Streaming itself is another active research topic: we considered the three most common approaches to tile-based streaming as well as a brief overview of viewport-dependent streaming. In particular, schemes that weigh the tiles by their viewing probability and importance in the projected \gls{fov} and maximize the overall expected \gls{qoe}, often including dynamic factors such as stalling and quality variations in the optimization, obtain the best performance. However, better \gls{fov} prediction is not the only way to improve streaming systems: additional options such as adaptive tiling schemes and \gls{svc} are also being investigated, as they can increase bandwidth efficiency and robustness in mobile streaming scenarios. Reinforcement learning-based schemes have recently been under their spotlight, as they can seamlessly integrate data from different sources in their prediction and optimize even complex \gls{qoe} functions in difficult scenarios with little design effort. Learning-based solutions provide higher accuracy and allow prediction for up to 10 seconds, a critical requirement to avoid stalling in buffer-based streaming systems.

Finally, network-level optimization to support omnidirectional streaming and \gls{vr} is another subject that is beginning to attract interest: the promises of 5G with regard to resource allocation and optimization, higher capacity, and edge and fog computing provide new interesting scenarios to simplify streaming systems and enable \gls{vr} over simple devices with limited battery and computing power.

Streaming techniques, along with all other aspects of omnidirectional video coding and evaluation, are rapidly converging towards machine learning as a general solution: the complexity of omnidirectional videos requires a level of context-awareness that is too complex for traditional analytical techniques. 
Furthermore, the trend in the field is towards \emph{joint} optimization, not considering each step of the process separately but optimizing them all at once, from projection and coding to streaming and quality evaluation. The first fully integrated models, incorporating historical data from other users, spatial and temporal features of the content, and past history for the specific user, are beginning to appear in the literature, although larger datasets with a varied population of viewers for proper evaluation are not available yet. Gaze tracking, which is more precise than head orientation tracking, is another possibility that is still largely unexplored due to the cost and complexity of the required experimental setup. However, the research related to several of the topics presented in this survey is still ongoing, and, given the fast update rate of communication technologies and the rapid growth of deep learning, we can expect the interest in the topic not to fade. In particular, \gls{vr} is central to the 5G paradigm, and innovations in each of the subjects we considered is needed to meet the high expectations.

\section*{References}

\bibliography{bibliography}

\printglossaries

\end{document}